%% file: main.tex
\documentclass[3p,onecolumn]{elsarticle}

\usepackage{xcolor}

\usepackage{amssymb}
\usepackage{amsmath}

\usepackage{algorithm}
\usepackage{pseudo}
\usepackage{booktabs}
\usepackage{multirow}
\usepackage{subcaption} 
\usepackage{diagbox}
\usepackage{xurl}
\usepackage{tikz}

\makeatletter
\@ifundefined{@textsuperscript}{}{%
  \let\els@orig@textsuperscript\@textsuperscript
  \def\@textsuperscript#1{{\check@mathfonts\els@orig@textsuperscript{#1}}}}
\makeatother

\usepackage[capitalise]{cleveref}

\journal{IFIP Performance 2026}

\newsavebox{\citationbox}
\usepackage{microtype}

\begin{document}

\begin{lrbox}{\citationbox}%
\begin{minipage}[t]{0.15\paperwidth}
\vspace{-0.5mm}%
\raggedleft BibTeX Citation:
\end{minipage}\hspace{5mm}%
\begin{minipage}[t]{0.8\paperwidth}\footnotesize
\vspace{-0mm}%
\begin{verbatim}
@article{LiPG26,
  author = {Zhuojin Li and Marco Paolieri and Leana Golubchik},
  title = {Partition-Aware Scheduling for Mobile Heterogeneous Inference Co-Execution},
  journal = {Perform. Evaluation}, volume = {}, number = {}, pages = {(to appear)},
  year = {2026}
}\end{verbatim}\end{minipage}%
\end{lrbox}

\AddToHook{shipout/foreground}{%
  \ifnum\value{page}=1
  \begin{tikzpicture}[remember picture,overlay]
    \node[anchor=north west,shift={(0,1.5mm)},minimum width=\paperwidth,
          minimum height=3cm,fill=black!5] at (current page.north west)
         {\usebox{\citationbox}};
  \end{tikzpicture}%
  \fi}
  
\begin{frontmatter}

\title{Partition-Aware Scheduling for Mobile Heterogeneous Inference Co-Execution}

\author[usc]{Zhuojin Li}
\ead{zhuojinl@usc.edu}
\author[usc]{Marco Paolieri}
\ead{paolieri@usc.edu}
\author[usc]{Leana Golubchik}
\ead{leana@usc.edu}
\affiliation[usc]{organization={University of Southern California},
            city={Los Angeles},
            state={CA},
            postcode={90089}, 
            country={USA}}

\input{sections/0-abstract}




\end{frontmatter}


\input{sections/1-introduction}

\input{sections/2-background}

\input{sections/3-motivation}

\input{sections/4-baseline}

\input{sections/5-method}

\input{sections/6-evaluation}

\input{sections/7-related-work}

\input{sections/8-conclusion}

\bibliographystyle{elsarticle-num} 
\bibliography{main}

\input{sections/appendix}

\end{document}

%% file: sections/0-abstract.tex
\begin{abstract}
Modern mobile inference runs on heterogeneous platforms combining mobile GPUs with multiple CPU core clusters.
Existing optimizations typically exploit either inter-operator parallelism, by assigning entire operators to CPU cores or to the GPU, or intra-operator parallelism, by partitioning each operator for CPU-GPU co-execution.
We consider these two forms of parallelism together, to improve inference latency of tasks that can be represented by a static DAG of operators with predefined input/output tensor shapes (e.g., CNNs or vision transformers). We define the problem of \emph{partition-aware DAG scheduling} for mobile heterogeneous inference, illustrating that the best strategy depends on the structure of the inference DAG, thus motivating a joint formulation capturing operator partition choices, device assignment, and execution order.
We propose an online iterative search framework, which decomposes large DAGs into stages, focuses search on critical operators, and uses latency predictors to estimate partitioned execution without exhaustive profiling.
Across representative mobile inference workloads, our approach achieves latency close to an offline solution while keeping scheduling overhead to a fraction of the model initialization cost, allowing platform-specific scheduling at deployment time.

\end{abstract}

%% file: sections/1-introduction.tex
\section{Introduction}\label{sec:introduction}

Deep learning has achieved significant breakthroughs across a wide range of applications, such as image understanding~\cite{krizhevsky2012imagenet}, speech recognition~\cite{hinton2012deep}, and augmented reality~\cite{lampropoulos2020enhancing}.
As neural networks increasingly support interactive applications, there is a growing need to execute inference directly on mobile platforms.
On-device inference improves offline availability, preserves user privacy by keeping data local, and enables low-latency responses without relying on cloud connectivity~\cite{chen2019deep}.
At the same time, state-of-the-art neural networks continue to exhibit increasing computational demands, requiring mobile platforms to incorporate more powerful compute resources for inference execution.
A modern mobile system typically contains multiple computing resources, including heterogeneous CPU clusters and a mobile GPU.
This hardware organization creates an important opportunity: rather than executing inference on a single accelerator, we can coordinate multiple compute resources at runtime to execute inference collaboratively.

For inference tasks where tensor shapes are fixed (e.g., convolutional neural networks and vision transformers), computation can be represented as a \textit{static} directed acyclic graph (DAG), where each node is a tensor operator (e.g., convolution) performing one computation, and each edge represents a data dependency.
This representation naturally leads to an optimization problem: assigning operators to heterogeneous devices and ordering their execution with respect to dependency constraints.
Classical heterogeneous DAG schedulers such as HEFT~\cite{topcuoglu2002performance} prioritize tasks and place each task on the device that gives the earliest finish time.
Such schedule-only methods exploit \textit{inter-operator parallelism}, where different operators can run concurrently on different devices (once their dependencies are satisfied).

However, DAG scheduling alone is limited in exploiting parallelism when the graph exposes little parallel structure.
For instance, many convolutional neural networks (CNNs) contain long chain-like regions, where most latency-dominant operators depend on the immediately preceding operator such that only one operator at a time is ready for computation.
In this case, even if the platform has idle compute resources, a schedule-only method cannot use them unless the operator itself can be partitioned.
For example, on ResNet16~\cite{he2016deep}, HEFT takes 6.8~ms and leaves the CPU idle for a significant fraction of the time because the model exposes limited inter-operator parallelism and the schedule is constrained by the critical path (\cref{sec:motivation_schedule_only}).

A complementary line of recent work uses CPU-GPU co-execution to split each operator across multiple devices~\cite{wang2018optic,jia2022codl,li2025epew}.
This approach exploits \textit{intra-operator parallelism}, where the workload of one operator is split into subtasks, and different devices compute different portions of the corresponding output.
Mobile platforms are well-suited to this approach because the CPU and GPU \textit{share unified memory}, allowing devices to access common tensor buffers without the overhead of extra copies; accordingly, in contrast to distributed computation on cloud GPUs~\cite{tpds22}, we do not model explicit CPU-GPU tensor-copy communication costs in this work because of our focus on mobile platforms with unified memory (see \cref{evaluation:discussion}).
Partition-only methods~\cite{li2025epew} typically make local decisions for each operator, i.e., choosing a workload split that minimizes the latency of that operator across compute units.
For chain-like graphs, such operator partitioning can create useful parallelism even when the DAG exposes few parallel operators.
In the ResNet16 example, this partition-only approach effectively utilizes the CPU resources and reduces latency to 4.9~ms.

However, partitioning every computationally expensive operator is not always beneficial.
In branch-heavy networks, such as Inception~\cite{szegedy2016rethinking} and HRNet~\cite{wang2020deep}, several operators may be ready at the same time.
These operators can have different device affinity, meaning that some run more efficiently on the GPU while others run competitively on CPU clusters.
In this setting, using all devices to accelerate one operator can delay another ready operator on the end-to-end critical path.
Thus, an operator partition that is locally beneficial can be globally harmful.
For example, on HRNet, partition-only execution takes $7.6$ ms, which is worse than schedule-only HEFT at $6.7$~ms, while jointly considering partitioning and scheduling reduces latency to $6.0$~ms.

These observations motivate \textit{partition-aware DAG scheduling}.
The scheduler needs to decide not only where to run each operator (and in what order), but also whether an operator should be split, how its workload should be divided, which devices should participate in the split, and how the resulting subtasks should be ordered under DAG dependencies.
This formulation generalizes both schedule-only and partition-only methods.
The key insight is that partitioning and scheduling are coupled in this formulation, as partitioning determines the tasks exposed to the scheduler, while scheduling determines whether using a partition improves end-to-end latency.

However, solving this joint problem is challenging because each partitioning decision changes the scheduling decision.
Even classical heterogeneous DAG scheduling is NP-hard~\cite{topcuoglu2002performance}.
Partition-aware scheduling adds another dimension of choices.
For each partitionable operator, the scheduler should further decide how to partition the operator and how to execute the partitions across available devices.
For a full DAG, these choices result in many possible task graphs expanded through node partitioning; each expanded graph must still be scheduled under device-availability and dependency constraints.
A mixed-integer optimizer such as Gurobi~\cite{gurobi} can solve small instances offline, but such solvers are unsuitable for deployment-time scheduling on mobile devices.

To make partition-aware DAG scheduling practical, we propose an iterative search framework designed for an online setting.
The framework starts from a feasible schedule and progressively improves it by exploring partitioning decisions that are likely to reduce end-to-end latency.
Three ideas make this search efficient.
First, \textit{staging} decomposes a large inference DAG into smaller sub-DAGs (or stages) and limits the search for the best schedule-partition to individual stages, thus reducing the optimization scope while preserving important dependencies at the boundary.
Second, \textit{criticality-aware sampling} focuses the search effort on operators that are likely to affect the current makespan, rather than spending equal effort on all nodes.
Third, \textit{latency prediction} estimates the cost of unseen partition plans, avoiding exhaustive on-device profiling for every candidate split.
Together, these techniques reduce the search space enough for online use, while retaining the main benefits of joint partitioning and scheduling.

This paper makes the following contributions:
\begin{itemize}

\item We empirically characterize the limitations of schedule-only and partition-only methods for mobile heterogeneous inference (\cref{sec:motivation}).
Chain-like models need intra-operator partitioning because they expose little DAG-level parallelism, while branch-heavy models require schedule-aware partitioning because locally beneficial splits can create global device contention.
These results motivate joint decisions over partitioning and scheduling, rather than relying on either approach alone.

\item We formulate partition-aware DAG scheduling as a unified optimization problem that captures operator partitioning, device assignment, and execution order (\cref{sec:formulation}).
This formulation provides a view of schedule-only and partition-only methods as restricted cases and gives an offline reference for evaluating practical algorithms.
Based on this formulation, we develop and analyze practical baselines including expanded-DAG scheduling and partition-aware list scheduling, showing that simply exposing additional parallel subtasks is insufficient to obtain the best end-to-end execution plan.

\item We design an iterative search framework for partition-aware scheduling (\cref{sec:iterative_search}) that can be used online, at deployment time.
To develop such an effective search for high-quality schedules, the framework uses staging to reduce scheduling scope, criticality-aware sampling to focus on operators that affect makespan, and latency prediction to evaluate unseen partition plans without exhaustive profiling.

\item We comprehensively evaluate the approach across 18 inference workloads and 4 mobile platforms (\cref{sec:evaluation}).
The iterative search achieves $1.01{\times}$--$1.04{\times}$ average normalized latency relative to Gurobi-based joint optimization (with a five-minute timeout for each stage) in a simulation setting and $1.00{\times}$--$1.04{\times}$ in real end-to-end measurements across devices.
The total deployment-time overhead, including latency prediction and scheduling, is only 37.4\% of model initialization time on average, justifying its practicality for on-device deployment.

\end{itemize}

%% file: sections/2-background.tex
\section{Background and System Model}\label{sec:background}

This section introduces the workload, hardware, and runtime assumptions used throughout the paper.
We first model neural-network inference as a DAG of tensor operators and define the heterogeneous devices available on a mobile platform (\cref{sec:background_inference}).
We then introduce workload partitioning, where a single operator is divided into subtasks that can execute collaboratively on multiple devices, and illustrate the partitioning strategies used for convolution operators (\cref{sec:background_partition}).
Lastly, we discuss why our scheduler is designed for deployment-time use during model initialization (\cref{sec:background_budget}).

\subsection{Mobile Heterogeneous Inference}\label{sec:background_inference}

A neural-network inference executes a trained model for a given input.
The model consists of a graph of \textit{operators}, where each operator performs one tensor computation, such as convolution, linear (matrix multiplication), or element-wise operators.
The input and output of each operator are \textit{tensors}, i.e., multi-dimensional arrays.
We represent an inference computation as a directed acyclic graph (DAG) $G=(V,E)$.
Each node $i\in V$ denotes an operator, and each edge $(i,j)\in E$ denotes a data dependency, indicating that operator $j$ can execute only after the required output of operator $i$ becomes available.
In this work, we consider fixed-shape inference workloads with a static operator graph and batch size of 1 (i.e., latency-oriented); batching and dynamic graphs are briefly discussed in \cref{evaluation:discussion}.

Modern mobile systems execute these workloads on heterogeneous compute resources.
Following the scheduling literature, we refer to each schedulable compute resource as a \textit{device}.
Our formulation and scheduling algorithms are defined over a device set $\mathcal{D}$, but for ease of presentation and to match the mobile platforms used in our evaluation, we instantiate $\mathcal{D}$ with three logical devices: GPU, CPU~(L), and CPU~(M).
Here, CPU (L) and CPU (M) denote the large-core and medium-core CPU clusters, respectively, as detailed in \cref{table:hardware_setup}.
This abstraction follows the clustered organization of modern mobile CPUs, in which cores within the same cluster have similar frequency and performance characteristics, while different clusters can exhibit substantially different energy efficiency and throughput.
Operators can exhibit different device affinity, meaning that an operator may execute faster on one device than another; this affinity depends on factors such as memory behavior, arithmetic intensity and device-specific operator implementation.
For example, large convolutions often benefit from GPU parallelism, while small or memory-bound operators may run competitively on CPU cores because they do not fully utilize the GPU.
Thus, the best device assignment can vary across operators and also depends on which other operators are ready to execute at the same time.
This consideration is crucial to our problem, as a device useful for accelerating one operator may be even more valuable for executing another \emph{ready operator} (i.e., an operator ready to execute) on the DAG critical path.

\begin{table}[t]
\centering
\scriptsize 
\begin{tabular}{l l l l l}
\toprule
\textbf{Platform} & \textbf{Processor} & \textbf{GPU} & \textbf{CPU (L)} & \textbf{CPU (M)} \\
\midrule
OnePlus 11 & Snapdragon 8 Gen 2 & Adreno 740 & $1{\times}$ 3.2 GHz Cortex-X3 & $2{\times}$ 2.8 GHz Cortex-A715 \\
\midrule
Motorola Edge Plus 2022 & Snapdragon 8 Gen 1 & Adreno 730 & $1{\times}$ 3.0 GHz Cortex-X2 & $2{\times}$ 2.5 GHz Cortex-A710 \\
\midrule
Pixel 5 & Snapdragon 765G & Adreno 620 & $1{\times}$ 2.4 GHz Kryo 475 & $1{\times}$ 2.2 GHz Kryo 475  \\
\midrule
Pixel 4 & Snapdragon 855 & Adreno 640 & $1{\times}$ 2.84 GHz Kryo 485  & $2{\times}$ 2.42 GHz Kryo 485 \\
\bottomrule
\end{tabular}
\caption{Mobile platforms used in the evaluation. CPU (L) and CPU (M) denote the large-core and medium-core CPU clusters exposed as logical scheduler devices in our experiments.}\label{table:hardware_setup}
\end{table}

The inference DAG exposes two forms of parallelism.
\textit{Inter-operator parallelism} executes multiple ready operators concurrently after their dependencies have been satisfied; for example, two independent branches of a network can run on different devices in parallel.
\textit{Intra-operator parallelism} divides one operator into multiple subtasks and executes them on multiple devices collaboratively, and we refer to this division as \textit{workload partitioning};
for example, a convolution can be partitioned so that the GPU computes part of the output tensor while a CPU cluster computes another part.

A key hardware property that makes this type of fine-grained co-execution practical on mobile platforms is \textit{unified memory}, where the CPU and GPU share the same physical memory.
Unlike systems with discrete GPUs, unified memory avoids large explicit data copies when CPU and GPU access the same tensors.
Moreover, following prior work~\cite{li2025epew}, our runtime uses OpenCL fine-grained shared virtual memory (SVM), which allows CPU and GPU to access shared allocations with hardware-supported cache coherence and avoids explicit data-mapping operations for maintaining coherence.
Therefore, partitioned operators can \textit{directly} read from and write to shared input and output tensors (using atomic operations).
Accordingly, in our scheduling model, we do \textit{not} include explicit CPU-GPU communication costs between operators.

Although unified memory removes the need to model explicit CPU-GPU communication costs, executing a schedule still requires lightweight coordination at dependency boundaries; e.g., a successor operator can start only after the required CPU and GPU work has completed.
Our runtime handles this coordination by inserting small GPU-side synchronization \textit{kernels} (i.e., small GPU programs) into the GPU command queue to update or wait on dependency progress shared with the CPU;
the full runtime implementation is detailed in the Appendix (\cref{sec:appendix:synchronization_details}).
These kernels perform little computation, but launching GPU work still incurs overhead because GPU commands are submitted by the CPU and executed asynchronously.
Accordingly, our scheduler accounts for GPU-side synchronization by adding a small constant to the GPU execution time;
this constant is set to 10~$\mu$s, based on the average overhead measured in our experiments.

\subsection{Workload Partitioning Strategies}\label{sec:background_partition}

Workload partitioning splits one operator into multiple subtasks, and each subtask computes a portion of the original operator's work and is assigned to one device.
The goal is to reduce operator latency (particularly for operators that contribute substantially to overall latency) by using otherwise idle devices.
In this paper, we focus on partitioning \textit{latency-dominant} operators (e.g., convolution and linear), i.e., those operators that dominate the latency of many mobile CNN workloads~\cite{li2024inference}.
Our framework is not limited to these operator types; in general, any operator can be treated as partitionable if its computation can be partitioned into subtasks.
Other operators, such as pooling and element-wise operators, are typically shorter and are treated as indivisible scheduling units.

We use standard convolution to illustrate the available partitioning strategies.
A linear layer can be handled similarly since it can be viewed as a convolution without a spatial window and its output features correspond to output channels.
Following common mobile ML framework layouts such as TensorFlow Lite (TFLite)~\cite{abadi2016tensorflow}, we describe tensor shape using height, width, and channel dimensions; since we consider batch size of 1, we omit the batch dimension.
For a convolution, the input tensor has shape $H_{\mathrm{in}}\times W_{\mathrm{in}}\times C_{\mathrm{in}}$, and the output tensor has shape $H_{\mathrm{out}}\times W_{\mathrm{out}}\times C_{\mathrm{out}}$.
The convolution uses a filter tensor $W$ of shape $K_h \times K_w \times C_{\mathrm{in}} \times C_{\mathrm{out}}$, where each output channel has one corresponding $K_h \times K_w \times C_{\mathrm{in}}$ filter.
Ignoring bias for simplicity, each output element is computed by applying a filter to a local spatial region of the input and accumulating the result across input channels: $Y[h,w,c_o] = \sum_{r=0}^{K_h-1} \sum_{s=0}^{K_w-1} \sum_{c_i=0}^{C_{\mathrm{in}}-1} X[h+r,w+s,c_i]\cdot W[r,s,c_i,c_o]$, for $0 \leq h < H_{\mathrm{out}}, 0 \leq w < W_{\mathrm{out}}, 0 \leq c_o < C_{\mathrm{out}}$.
This formula reveals three natural ways to split the computation: output channels, spatial positions, and input channels.

\textit{Output-Channel Partitioning.}
Output-channel partitioning splits the $C_{\mathrm{out}}$ dimension.
Each device computes a subset of output channels using the corresponding filters.
Because different output channels are independent, devices compute disjoint regions of the output tensor.
Under unified memory, these regions can be written directly into the shared output buffer using atomic operations, without additional copy or aggregation.
Output-channel partitioning therefore provides a low-overhead partitioning strategy for convolution and linear operators.

\textit{Spatial Partitioning.}
Spatial partitioning splits the output spatial dimensions, such as the width dimension $W_{\mathrm{out}}$.
Each device computes a contiguous range of output positions.
For $1\times1$ convolutions, the split is naturally independent because each output position depends only on the corresponding input position.
For larger kernels, however, output elements near a partition boundary require overlapping neighboring input values outside the assigned output range.
Unified memory allows these boundary values to be read without explicit CPU-GPU copies.

\textit{Input-Channel Partitioning.}
Input-channel partitioning splits the $C_{\mathrm{in}}$ dimension.
Unlike output-channel and spatial partitioning, this split does not directly produce independent final outputs.
The devices therefore accumulate their partial sums into the shared output tensor using atomic updates.
Under unified memory, the overhead of dispatching a separate aggregation kernel can be avoided by modifying the compute kernel to update the final output tensor directly.

\subsection{Online Scheduling Budget}\label{sec:background_budget}

The previous subsections define the decisions available to the scheduler; it can exploit inter-operator parallelism by running ready operators on different devices, and intra-operator parallelism by partitioning one operator across multiple devices.
However, exploring these decisions introduces scheduling overhead.
In this paper, we use \textit{online} scheduling to refer to scheduling determined during model initialization, rather than during an expensive offline tuning phase.
Similar to model initialization, the scheduling decisions do not need to be recomputed before every inference request.
For a fixed model and fixed input shape, the runtime can determine the execution plan once and reuse it for subsequent inferences.
Thus, scheduling overhead is amortized across many requests.
From this perspective, a scheduler does not need to be as fast as a single inference, but it should add only modest overhead relative to model initialization.
For example, on OnePlus 11, the TFLite benchmark reports 1048~ms to initialize Inception-v3, including weight loading, tensor allocation, graph compilation, and device-specific preparation.
A practical online scheduler should execute during this initialization.
This requirement prevents the scheduler from exhaustively profiling every partition choice on the device, or solving a large global optimization problem with an offline solver.

%% file: sections/3-motivation.tex
\section{Motivation: Why Partitioning and Scheduling Should Be Joint}\label{sec:motivation}

As noted above, mobile DNN inference exposes two forms of parallelism:
(1) the inference DAG can contain branches that can run concurrently on different devices, i.e., inter-operator parallelism;
(2) a latency-dominant operator (e.g., convolution) can be partitioned into subtasks that run collaboratively on multiple devices, i.e., intra-operator parallelism.
This section considers (empirically) three representative neural networks to illustrate that exploiting only one form of parallelism can be suboptimal (\cref{sec:motivation_schedule_only,sec:motivation_partition_only}), motivating the need to jointly optimize partitioning and scheduling (\cref{sec:motivation_joint}).

\begin{figure}[t]
	\centering
	\begin{subfigure}[b]{.24\linewidth}
		\centering
		\includegraphics[width=\linewidth]{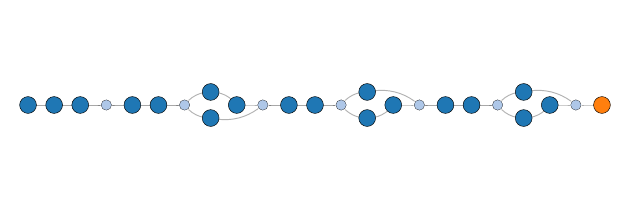}
		\caption{Chain-dominated: ResNet16}\label{fig:architecture_resent}
	\end{subfigure}
	\begin{subfigure}[b]{.37\linewidth}
		\centering
		\includegraphics[width=\linewidth]{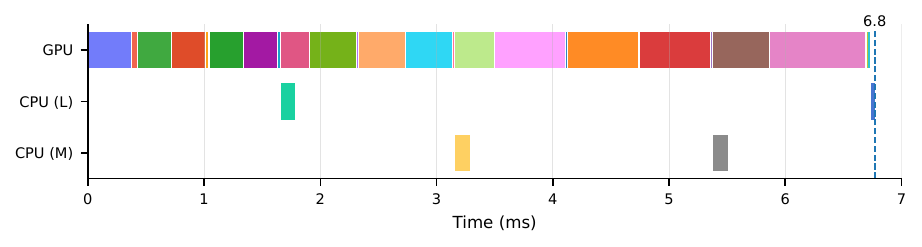}
		\caption{Schedule-only: ResNet16}\label{fig:resnet16_schedule_only}
	\end{subfigure}
	\begin{subfigure}[b]{.37\linewidth}
		\centering
		\includegraphics[width=\linewidth]{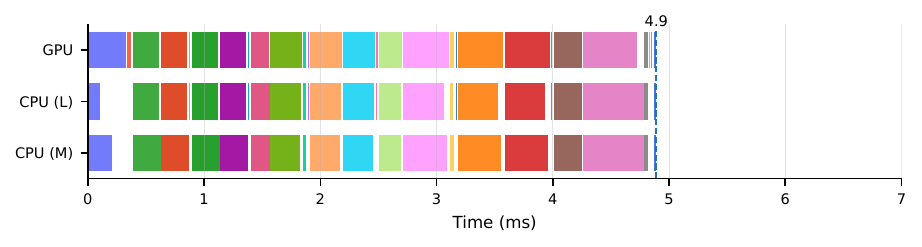}
		\caption{Partition-only: ResNet16}\label{fig:resnet16_partition_only}
	\end{subfigure}
	\caption{Chain-dominated ResNet16 on OnePlus 11 phone. Schedule-only execution leaves CPU clusters mostly idle because few operators are ready at the same time. Partitioning operators creates intra-operator parallelism.}\label{fig:motivation_chain_dominated}
\end{figure}

\begin{figure}[t]
	\centering
	\begin{subfigure}[b]{.48\linewidth}
		\centering
		\includegraphics[width=\linewidth]{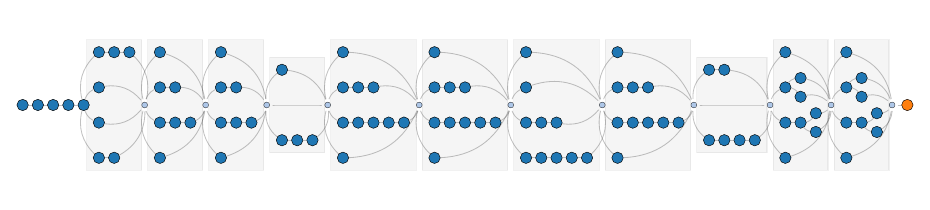}
		\caption{Block-structured graph: Inception-v3}\label{fig:architecture_inceptionv3}
	\end{subfigure}
	\begin{subfigure}[b]{.48\linewidth}
		\centering
		\includegraphics[width=\linewidth]{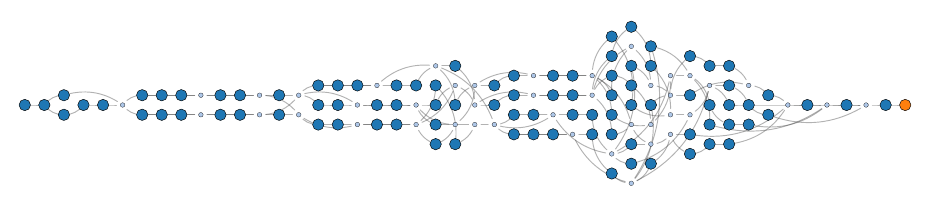}
		\caption{Branch-heavy graph: HRNet}\label{fig:architecture_hrnet}
	\end{subfigure}
	\begin{subfigure}[b]{.49\linewidth}
		\centering
		\includegraphics[width=\linewidth]{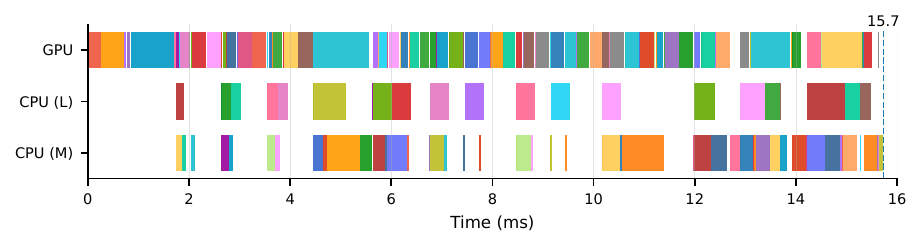}
		\caption{Schedule-only: Inception-v3}\label{fig:inceptionv3_schedule_only}
	\end{subfigure}
	\begin{subfigure}[b]{.49\linewidth}
		\centering
		\includegraphics[width=\linewidth]{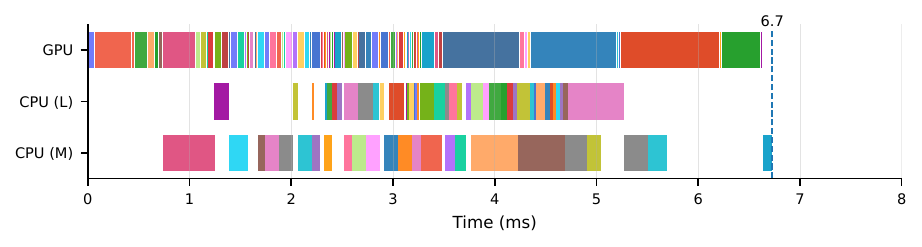}
		\caption{Schedule-only: HRNet}\label{fig:hrnet_schedule_only}
	\end{subfigure}

	\begin{subfigure}[b]{.49\linewidth}
		\centering
		\includegraphics[width=\linewidth]{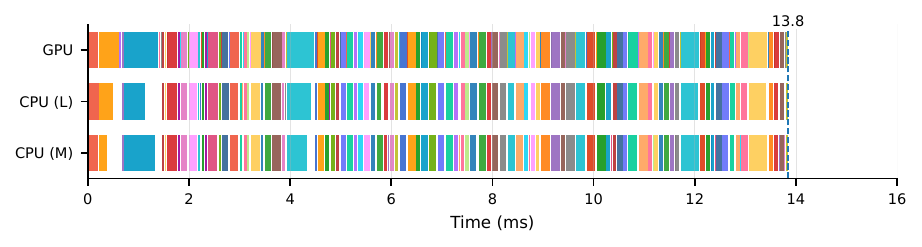}
		\caption{Partition-only: Inception-v3}\label{fig:inceptionv3_partition_only}
	\end{subfigure}
	\begin{subfigure}[b]{.49\linewidth}
		\centering
		\includegraphics[width=\linewidth]{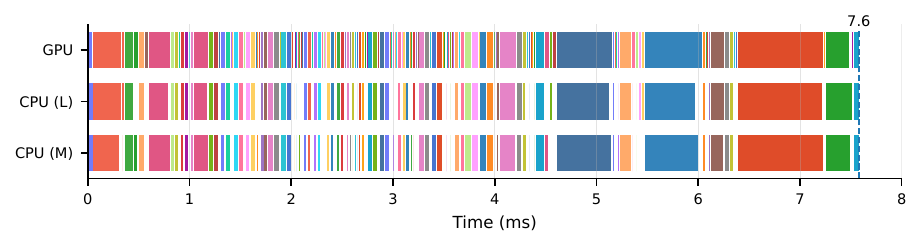}
		\caption{Partition-only: HRNet}\label{fig:hrnet_partition_only}
	\end{subfigure}

	\begin{subfigure}[b]{.49\linewidth}
		\centering
		\includegraphics[width=\linewidth]{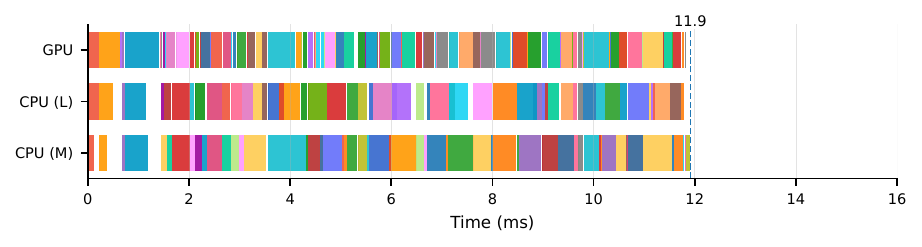}
		\caption{Partition and schedule: Inception-v3}\label{fig:inceptionv3_partition_schedule}
	\end{subfigure}
	\begin{subfigure}[b]{.49\linewidth}
		\centering
		\includegraphics[width=\linewidth]{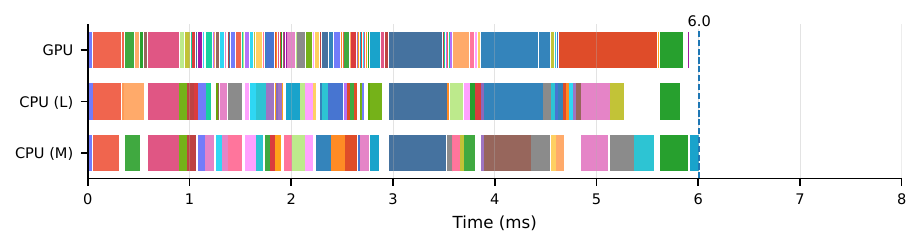}
		\caption{Partition and schedule: HRNet}\label{fig:hrnet_partition_schedule}
	\end{subfigure}

	\caption{Branch-structured models on OnePlus 11 phone. Inception-v3 has limited branch-level parallelism, so partitioning can use idle devices. HRNet exposes more ready operators, so partition-only execution can block useful inter-operator parallelism. Joint partitioning and scheduling balance these effects.}\label{fig:motivation_branch_heavy}
\end{figure}

\subsection{Representative Model Structures}

We study three representative models: ResNet16~\cite{he2016deep}, Inception-v3~\cite{szegedy2016rethinking}, and HRNet~\cite{wang2020deep}; they exhibit different degrees of intra- and inter-operator parallelism.
ResNet16 is chain-dominated: as shown in \cref{fig:architecture_resent}, its latency-dominant convolutions are primarily organized along a sequential critical path, and at most points in the execution, only one operator is ready to execute.
Therefore, the main optimization opportunity is to reduce the latency of each critical operator by partitioning each operator across multiple devices.
Inception-v3 has a block-structured DAG; as shown in \cref{fig:architecture_inceptionv3}, each Inception block (in a shaded rectangular region) creates several parallel branches after a split point and later merges them through concatenation.
This structure exposes more inter-operator parallelism than ResNet16, but the parallelism is still limited by block boundaries and by imbalance across branches.
Thus, Inception-v3 can benefit from both branch-level scheduling and operator partitioning.
HRNet is more branch-heavy; as shown in \cref{fig:architecture_hrnet}, it maintains multiple parallel branches through a large portion of the network.
This structure creates a wide middle region in which many operators can be ready at the same time.
In this setting, aggressively partitioning every operator can be ineffective because devices used for one partitioned operator may be more suitable for other ready operators from parallel branches.

\cref{fig:motivation_chain_dominated,fig:motivation_branch_heavy} also show execution timelines of different methods on three devices.
Each colored block represents the execution interval of an operator or an operator partition (i.e., a subtask).
In schedule-only execution, each operator is assigned to one device.
In partition-only execution, every partitionable operator is split across all three devices; blocks with the same color correspond to partitions of the same original operator.
In joint partition-aware scheduling (\cref{fig:inceptionv3_partition_schedule,fig:hrnet_partition_schedule}), the scheduler decides both how each operator is partitioned and how the resulting subtasks should be placed (and ordered) in the DAG schedule.

\subsection{Schedule-Only Execution Misses Intra-Operator Parallelism}\label{sec:motivation_schedule_only}

Schedule-only execution treats each operator as an indivisible task.
A representative method is HEFT~\cite{topcuoglu2002performance}, which first prioritizes operators based on an estimate of their remaining critical-path cost and then assigns each operator to the device that gives the earliest finish time (as detailed in \cref{sec:baseline_schedule_only}). 
Such methods are effective when the DAG contains enough ready operators to keep multiple devices busy.
However, they cannot create additional parallel work when the DAG exposes little inter-operator parallelism.
This limitation is clear for chain-dominated ResNet16.
As shown in \cref{fig:resnet16_schedule_only}, HEFT takes 6.8~ms, where the GPU executes most latency-dominant operators, while CPU (L) and CPU (M) remain almost idle (with utilization below 4\%);
this is mainly because most operators depend on the immediately preceding operator, so there are few ready operators for the CPU clusters to execute.
As a result, a schedule-only method cannot use idle devices unless there is another ready operator.
On the other hand, partitioning addresses this limitation by creating parallel work inside each operator.
As shown in \cref{fig:resnet16_partition_only}, partitioned execution splits latency-dominant operators across GPU and CPU clusters, which increases the utilization of all devices to more than 85\% and reduces latency to 4.9~ms.

\subsection{Partition-Only Execution Misses Inter-Operator Parallelism}\label{sec:motivation_partition_only}

Partition-only execution typically optimizes each partitionable operator locally.
A common objective is to divide one operator across devices so that the participating devices finish their partitions at similar times~\cite{jia2022codl, li2025epew}.
This local balancing can reduce the latency of an individual operator, particularly when some devices would otherwise be idle.
Inception-v3 illustrates a case where partitioning remains useful.
This model contains branch-level parallelism, so schedule-only execution can place independent branch operators on different devices, as shown in \cref{fig:inceptionv3_schedule_only}.
However, the branches inside each block are not always sufficient to fully occupy the slower CPU clusters; the resulting CPU utilizations are only 43\% for CPU (L) and 52\% for CPU (M).
Instead, partition-only execution (\cref{fig:inceptionv3_partition_only}) can utilize these idle devices, reducing latency to 13.8~ms compared with 15.7~ms for schedule-only execution.

However, the local balancing objective can become ineffective when the DAG already exposes abundant inter-operator parallelism.
For example, HRNet contains a wide middle region with multiple parallel branches, where many operators can be ready at the same time.
Schedule-only execution can exploit this structure; much of the timeline in \cref{fig:hrnet_schedule_only} is fully utilized within the 2.5--5~ms window, where independent branch operators occupy all the devices.
On the other hand, partition-only execution (\cref{fig:hrnet_partition_only}) locally partitions each operator across all devices, ignoring whether the devices used for one operator partition could have been used more effectively by other ready operators.
As a result, partition-only execution takes 7.6~ms, which is worse than the 6.7~ms achieved by schedule-only execution.
Joint partition-aware scheduling avoids this problem by partitioning an operator only when its intra-operator speedup outweighs the benefit of using the same device for other ready operators.
For HRNet, jointly optimizing partitioning and scheduling reduces latency to 6.0~ms, as shown in \cref{fig:hrnet_partition_schedule}.

\subsection{Motivation for Joint Partition-Aware Scheduling}\label{sec:motivation_joint}

The examples above illustrate that the effectiveness of an execution strategy is closely tied to model topology.
In chain-dominated regions, such as ResNet16, execution is primarily constrained by a sequential critical path, so reducing the latency of individual operators through intra-operator partitioning is highly effective.
In branch-heavy regions, such as HRNet, many operators can be ready at the same time, and these operators compete for heterogeneous devices;
in this case, partitioning one operator across all devices can reduce its local latency while increasing the end-to-end makespan by delaying other ready operators.
Therefore, the scheduler must evaluate each partitioning decision in the context of other ready operators in the graph.

These observations motivate \textit{joint partition-aware DAG scheduling}.
Instead of choosing between schedule-only and partition-only execution, the scheduler should jointly decide which operators to partition, how to divide their workload, which devices should execute the resulting subtasks, and when those subtasks should run.
The guiding principle is to partition operators when device idleness limits performance, while taking advantage of inter-operator parallelism when ready operators can already keep the devices busy.

%% file: sections/4-baseline.tex
\section{Problem Formulation and Baselines}\label{sec:formulation}

This section formalizes partition-aware DAG scheduling (\cref{sec:baseline_formulation}) and defines the baseline methods used in our evaluation.
The formulation decomposes the problem into two decisions: \textit{partitioning}, which determines the partition strategy and the workload division for each operator, and \textit{scheduling}, which assigns the resulting subtasks to devices and orders their execution while respecting DAG dependencies.
From this perspective, existing approaches can be viewed as restricted versions of the joint problem:
schedule-only methods (\cref{sec:baseline_schedule_only}) optimize device placement and execution order but treat operators as indivisible,
while partition-only methods (\cref{sec:baseline_partition_only}) optimize local operator partitioning but omit graph-level scheduling.

In addition to these existing baselines, we propose two simple partition-aware baselines to evaluate whether straightforward combinations of partitioning and scheduling are sufficient.
The first, \emph{expanded-DAG scheduling} (\cref{sec:baseline_expanded_dag}), fixes a partition plan for each operator, expands the original DAG into subtasks, and then applies a conventional DAG scheduler.
The second, \emph{partition-aware list scheduling} (\cref{sec:baseline_partition_aware_heuristics}), extends schedule-only list scheduling by greedily choosing a partition plan when each operator is scheduled.
Finally, the full joint formulation provides an offline reference problem that can be solved by an optimizer (e.g., Gurobi) for small graphs; this reference allows us to evaluate how close practical methods are to state-of-the-art optimization with a long timeout (5 minutes in our experiments).

\begin{table}[t]
\centering
\footnotesize
\begin{tabular}{ll}
\toprule
Symbol & Description \\
\midrule

$G=(V,E)$ & Inference DAG with operators $V$ and dependencies $E$ \\
$\mathrm{pred}(i),\mathrm{succ}(i)$ & Predecessor and successor sets of operator $i$ \\
$\mathcal{D}$ & Set of devices, e.g., GPU and CPU core clusters \\
$V_p$ & Set of partitionable operators \\
$\mathcal{A}_i$ & Candidate partition strategies for operator $i$ \\
$a_i$ & Selected partition strategy of operator $i$, e.g., none, output-channel, input-channel, or spatial \\
$K_i$ & Maximum number of subtasks for operator $i$ \\
$U_{i,a}$ & Integer workload size of operator $i$ under strategy $a$ \\
$\mathbf{u}_i$ & Workload division vector $(u_{i1},\ldots,u_{iK_i})$, where $u_{iq}$ is workload size for subtask $(i,q)$ \\
$\pi_i$ & Specific partition plan of operator $i$, $\pi_i=(a_i,\mathbf{u}_i)$ \\
$\Pi_i$ & Set of candidate partition plans for operator $i$ \\
$\boldsymbol{\pi}$ & Partition plans for all operators, $\boldsymbol{\pi}=\{\pi_i\}_{i\in V}$ \\
$\mathbf{d}^{\mathrm{OSPD}}$ & One-subtask-per-device mapping used for local partition decisions \\
$\mathbf{d}_i$ & Device assignment vector $(d_{i1},\ldots,d_{iK_i})$, where $d_{iq}$ is the assignment for subtask $(i,q)$ \\
$\hat{\tau}_{i,a,d}(u)$ & Estimated latency for workload $u$ on device $d$ under strategy $a$ \\
$s_{iq}, e_{iq}$ & Start and end times of subtask $(i,q)$ \\
$\ell_{iq}$ & Duration of subtask $(i,q)$, $\ell_{iq}=e_{iq}-s_{iq}$ \\
$C_i$ & Completion time of operator $i$ \\
$T$ & End-to-end makespan \\

$w_{i,d}$ & Unpartitioned execution time of operator $i$ on device $d$ \\
$\bar{w}_i$ & Average unpartitioned execution time of operator $i$ across devices \\

$S$ & A constructed schedule, including subtask placement and execution order \\
$T(S)$ & Makespan of schedule $S$ \\
$\mathrm{rank}_u(i)$ & HEFT upward rank of operator $i$ \\
$\mathrm{EST}(\cdot), \mathrm{EFT}(\cdot)$ & Earliest start time and earliest finish time used by list scheduling \\

\bottomrule
\end{tabular}
\caption{Notation used in the optimization model and baseline definitions.}\label{table:notations}
\end{table}

\subsection{Formulation: Partition-Aware DAG Scheduling}\label{sec:baseline_formulation}

\textit{DAG and Devices.}
We model an inference computation as a directed acyclic graph (DAG) $G=(V,E)$, where each node $i \in V$ is an operator and each directed edge $(i,j) \in E$ denotes that operator $j$ depends on the output of operator $i$.
The target mobile platform contains a set of heterogeneous devices $\mathcal{D}$, including mobile GPU and CPU core clusters.
Each device can execute at most one scheduled task or subtask at a time.
Because the CPU and GPU share unified memory, we do not explicitly model data-copy communication cost between dependent operators.

\textit{Partition Strategies.}
Unlike traditional DAG scheduling, where each operator is treated as an indivisible task, our formulation allows a subset $V_p\subseteq V$ of operators  to be partitioned.
In this paper, partitionable operators primarily include convolution and linear operators, which typically dominate end-to-end latency in mobile inference workloads~\cite{li2024inference}.
For each partitionable operator $i$, we consider a set of candidate partition strategies
$A_i = \{\mathrm{none}\} \cup A_i^{\mathrm{part}}, A_i^{\mathrm{part}} \subseteq \{c_{\mathrm{out}},c_{\mathrm{in}},\mathrm{spatial}\}$.
Here, $\texttt{none}$ denotes unpartitioned execution,
$c_{\mathrm{out}}$ denotes output-channel partitioning,
$c_{\mathrm{in}}$ denotes input-channel partitioning, and
$\texttt{spatial}$ denotes spatial partitioning (as described in \cref{sec:background_partition}).
For non-partitionable operators, we set $\mathcal{A}_i=\{\texttt{none}\}$.

\textit{Partition Plan and Scheduling Decision.}
For a strategy $a \in \mathcal{A}_i$, let $U_{i,a} \in \mathbb{Z}_{>0}$ denote the (integer) workload size of operator $i$ under that strategy;
we model workload size as an integer to reflect actual partition strategies (see \cref{sec:background_partition}).
For example, $U_{i,c_{\mathrm{out}}}$ is the number of output channels of operator $i$ when output-channel partitioning is used.
A specific partition plan for operator $i$ is $\pi_i=(a_i,\mathbf{u}_i)$, where $a_i \in \mathcal{A}_i$ is the selected strategy, and $\mathbf{u}_i=(u_{i1},u_{i2},\ldots,u_{iK_i})$ denotes the workload division, satisfying $u_{iq}\in\mathbb{Z}_{\ge 0}$ and $\sum_{q=1}^{K_i}u_{iq} = U_{i,a_i}$.
Subtasks with $u_{iq}=0$ are inactive and are omitted from the resulting task graph and scheduling decisions.
We allow zero-work entries so that the formulation supports partitioning into \textit{up to} $K_i$ subtasks.
For each partitionable operator $i$, $\Pi_i$ represents the set of feasible partition plans, i.e., combinations of partition strategy and workload division into at most $K_i=|\mathcal{D}|$ subtasks.
For every operator $i\in V$, the unpartitioned strategy corresponds to the plan $\pi_i^{\mathrm{none}} =(\mathrm{none},(U_{i,\mathrm{none}},0,\ldots,0))$.
For a non-partitionable operator $i\notin V_p$, we set $\Pi_i=\{\pi_i^{\mathrm{none}}\}$.
Given a specific partition plan $\pi_i=(a_i,\mathbf{u}_i)$, the scheduler determines device placement and the execution order of the resulting subtasks.
Let $\mathbf{d}_i=(d_{i1},d_{i2},\ldots,d_{iK_i})$ denote the device assignment vector for operator $i$, where $d_{iq}\in\mathcal{D}$ is the device assigned to subtask $(i, q)$.
For a given operator, we require its active subtasks to be assigned to distinct devices; i.e., $d_{iq}\neq d_{ir}$ for $q\neq r$ whenever $u_{iq}>0$ and $u_{ir}>0$.
The scheduler also orders execution on each device by assigning start and end times $s_{iq}$ and $e_{iq}$, subject to device availability and DAG dependencies.
As a result, the final makespan is determined by both the partition plans for all operators and the scheduling decisions that assign tasks to devices and order their execution.

\textit{Execution Time.}
For a specific partition plan $\pi_i=(a_i,\mathbf{u}_i)$ and device assignment $\mathbf{d}_i$, the estimated latency of subtask $q$ is denoted as $\hat{\tau}_{i,a_i,d_{iq}}(u_{iq}) \geq 0$.
Notably, the estimate includes the device execution time under the selected strategy; for GPU subtasks, it also includes the small constant that accounts for synchronization kernel-dispatch overhead, as described in \cref{sec:background_inference}.
The end time of a subtask is $e_{iq} = s_{iq} + \hat{\tau}_{i,a_i,d_{iq}}(u_{iq})$.
The completion time of the entire operator $i$ is determined by the slowest subtask $C_i = \max_{q:u_{iq}>0} e_{iq}$.
Our formulation enforces non-overlapping execution on each device; i.e., each device can execute at most one subtask at a time.
Therefore, if two subtasks are assigned to the same device, their execution intervals cannot overlap: $ [s_{iq},e_{iq}) \cap [s_{jr},e_{jr}) = \emptyset$ if $d_{iq}=d_{jr}$ and $(i,q)\neq (j,r)$.

\textit{Full-Barrier Dependency.}
We use a full-barrier dependency model, where a successor operator can start only after all subtasks of each predecessor have completed.
For each edge $(i,j)\in E$, we enforce $s_{jq} \ge C_i$ for $q\in\{1,\ldots,K_j\}$.
This captures the requirement that the successor operator consumes the full output of each predecessor.

\textit{Number of Subtasks.}
In this paper, we set $K_i=|\mathcal{D}|$, indicating that each operator creates at most one subtask per device.
Under the full-barrier dependency model, creating more subtasks than available devices would require some subtasks of the same operator to execute serially on the same device, which rarely increases parallelism in our setting.
We validate this assumption through an ablation study in~\cref{sec:evaluation_ablation}.

\textit{Objective.}
The joint problem optimizes both partitioning plans $\pi_i=(a_i,\mathbf{u}_i)$ and scheduling decisions $(\mathbf{d}_i,s_{iq},e_{iq})$,
with the objective of minimizing the end-to-end inference makespan: $\min T \;\text{s.t.}\; T \ge C_i,\ \forall i \in V$.

This formulation captures intra-operator parallelism through partitioning and inter-operator parallelism through scheduling.
We implement this joint optimization problem in Gurobi, where partition-plan selection, device assignment, and non-overlap constraints are encoded using binary selection and ordering variables.
Because the search space grows rapidly with the number of partitionable operators and candidate partition plans, the optimizer is used only as an offline baseline (with a five-minute solution timeout for each stage).

\subsection{Schedule-Only Heuristics}\label{sec:baseline_schedule_only}

Schedule-only methods exploit inter-operator parallelism by assigning operators to different devices, but each operator remains indivisible.
This approach corresponds to fixing the partition plan of every operator~$i$ to $\pi_i^{\mathrm{none}} = \left( \texttt{none}, (U_{i,\texttt{none}},0,\ldots,0) \right)$.
The scheduler then chooses the device placement and execution order of operators.
Under this restriction, the problem reduces to classical heterogeneous DAG scheduling.

We use HEFT~\cite{topcuoglu2002performance} as a representative schedule-only heuristic.
For an operator $i$, let $w_{i,d} = \hat{\tau}_{i,\texttt{none},d}(U_{i,\texttt{none}})$ denote the execution time of operator $i$ on device $d$.
HEFT first computes an average execution time across devices $\bar{w}_i = \frac{1}{|\mathcal{D}|}\sum_{d\in\mathcal{D}} w_{i,d}$.
The original HEFT formulation also includes an average communication cost between dependent tasks; we omit this communication term because CPU and GPU share unified memory.
The upward rank is defined as $\mathrm{rank}_u(i) = \bar{w}_i + \max_{j\in \mathrm{succ}(i)} \mathrm{rank}_u(j)$, where $\mathrm{succ}(i)=\{j:(i,j)\in E\}$ is the set of successors of operator $i$ (we set the $\max$ to 0 when $i$ has no successors).
HEFT schedules operators in decreasing order of $\mathrm{rank}_u(i)$.
When operator $i$ is selected, HEFT assigns it to the device that gives the earliest finish time: $d_i^\star = \arg\min_{d\in\mathcal{D}} \mathrm{EFT}(i,d)$, where $\mathrm{EFT}(i,d)=\mathrm{EST}(i,d)+w_{i,d}$.
Here, $\mathrm{EST}(i,d)$ is the earliest time at which operator $i$ can start on device $d$, considering both predecessor completion times and the current availability of device $d$.
In summary, HEFT first prioritizes operators that are important to the remaining critical path, and then greedily assigns each operator to the device that achieves the earliest finish time.

We also evaluate several HEFT-like list-scheduling heuristics.
Lookahead-HEFT~\cite{bittencourt2010dag} improves the greedy device selection by considering how assigning the current operator to a device affects its immediate successors.
PEFT~\cite{arabnejad2013list} uses an optimistic cost table to estimate the downstream scheduling cost, so that device selection accounts for future critical-path effects rather than only the current earliest finish time.
LDCP~\cite{daoud2008high} prioritizes operators using a longest dynamic critical-path criterion, so operators more likely to delay the final makespan are scheduled earlier.
CEFT~\cite{khan2012scheduling} uses constrained critical path information to guide operator prioritization and device selection.
PSLS~\cite{zhao2019list} uses a pre-scheduling phase to obtain global scheduling guidance before applying list scheduling.
These methods improve different parts of the list-scheduling procedure, but they all keep operators indivisible.
We also include a schedule-only Gurobi baseline for small DAGs.
%
This baseline optimizes placement and ordering of indivisible operators, allowing us to separate the benefit of optimal graph-level scheduling.

\subsection{Partition-Only Co-Execution}\label{sec:baseline_partition_only}

Partition-only methods exploit intra-operator parallelism but do not perform graph-level scheduling.
Existing work~\cite{jia2022codl,li2025epew} typically assumes a fixed co-execution device placement, which we refer to as the \textit{one-subtask-per-device} mapping, abbreviated as OSPD.
Under OSPD, each subtask of an operator is mapped to a different device:
$\mathbf{d}^{\mathrm{OSPD}} = (d^{\mathrm{OSPD}}_1,\ldots,d^{\mathrm{OSPD}}_{|\mathcal{D}|})$,
with $\cup_{q=1}^{|\mathcal{D}|}\,\{d^{\mathrm{OSPD}}_q\} = \mathcal{D}$.
Given this fixed device placement, partition-only co-execution chooses the partition plan that locally minimizes the overall latency of each partitionable operator:
$\pi_i^{\mathrm{Local}} = \arg\min_{\pi_i=(a_i,\mathbf{u}_i)\in\Pi_i}  \max_{q=1,\ldots,K_i} \hat{\tau}_{i,a_i,d^{\mathrm{OSPD}}_q}(u_{iq})$.
The graph then follows the original topological execution order without optimizing operator order or device placement using graph-level information.

A partition-only method can significantly accelerate heavy operators by dividing their work across CPU and GPU.
However, this decision is made locally for each operator and does not consider whether keeping a device available would better serve other ready operators.
For example, splitting the current operator across both CPU and GPU can minimize the operator latency locally, but it can also delay another ready operator that has stronger affinity to one of the devices.
Therefore, partition-only co-execution can be effective for chain-dominated DAGs, but it can underutilize graph-level parallelism in branch-heavy DAGs.

\subsection{Proposed Baseline: Expanded-DAG Scheduling}\label{sec:baseline_expanded_dag}

Expanded-DAG scheduling is a simple attempt to combine partitioning with existing DAG schedulers.
It decouples the problem into two stages.
First, each partitionable operator is assigned a specific partition plan $\pi_i=(a_i,\mathbf{u}_i)$.
Second, the original DAG is expanded into subtasks (based on the partition plan) preserving the full-barrier dependency. A conventional schedule-only heuristic or optimizer then schedules the expanded graph.
We consider two fixed partition rules.

\textit{Equal-Size Expansion.}
The equal-size rule divides (integral size) workload evenly (based on the partitioning strategy used) across subtasks, resulting in an equal amount of work per subtask.
For each strategy $a \in \mathcal{A}_i$, the workload division follows $u_{iq}^{\mathrm{equal}}(a) \approx \frac{U_{i,a}}{K_i}$.
We then choose the strategy whose equal-size partition has the smallest latency under the OSPD placement:
$a_i^{\mathrm{equal}} = \arg\min_{a\in\mathcal{A}_i} \left( \max_q \hat{\tau}_{i,a,d^{\mathrm{OSPD}}_q} \left( u_{iq}^{\mathrm{equal}}(a) \right)  \right)$.
The resulting fixed partition plan is $\pi_i^{\mathrm{equal}} = \left( a_i^{\mathrm{equal}}, \mathbf{u}_i^{\mathrm{equal}}(a_i^{\mathrm{equal}}) \right)$.

\textit{Min-Local-Latency Expansion.}
Unlike equal-size expansion, which fixes an equal amount of work distributed across devices, this rule searches over workload divisions to choose the one that minimizes the latency of the operator locally.
The min-local-latency rule chooses the partition plan that locally minimizes the latency of operator $i$ under the OSPD placement:
$\pi_i^{\mathrm{Local}} = \arg\min_{\pi_i=(a_i,\mathbf{u}_i)\in\Pi_i}  \max_{q=1,\ldots,K_i} \hat{\tau}_{i,a_i,d^{\mathrm{OSPD}}_q}(u_{iq})$.
This rule follows the same local objective as partition-only co-execution.

Both equal-size expansion and min-local-latency expansion fix the partition plan of each operator; graph-level scheduling is performed after the expanded graph is constructed.
Expanded-DAG scheduling evaluates candidate solutions using a decoupled design, where partitioning is decided locally first, and scheduling is performed afterward.
However, because the partition plans are fixed before graph-level scheduling, they do not account for how devices will be assigned to other ready operators.
As a result, a partition that is locally beneficial for one operator may still be globally suboptimal.

\subsection{Proposed Baseline: Partition-Aware List Scheduling}\label{sec:baseline_partition_aware_heuristics}

Partition-aware list scheduling incorporates partitioning decisions during scheduling.
Unlike expanded-DAG scheduling, it does not fix all partition plans before scheduling begins.
Instead, when an operator is selected by the list scheduler, the heuristic greedily chooses both a partition plan and a device placement for that operator.
We present a partition-aware list scheduling extension of HEFT as an example.

\textit{Partition-Aware Priority.}
Standard HEFT computes an upward rank using the average unpartitioned execution time of each operator across devices.
However, this cost estimate does not reflect the fact that a partitionable operator may execute faster when its work is divided across multiple devices.
To account for partitionable execution, we estimate the cost of each operator using its minimized local latency under the OSPD mapping: $\bar{w}^{\mathrm{PA}}_i = \min_{\pi_i=(a_i,\mathbf{u}_i)\in\Pi_i}  \max_q \hat{\tau}_{i,a_i,d^{\mathrm{OSPD}}_q}(u_{iq})$.
This is the same local cost used by the partition-only baseline.
Using this effective cost, the partition-aware (PA) upward rank follows the same recursive structure as HEFT:
$\mathrm{rank}^{\mathrm{PA}}_u(i) = \bar{w}^{\mathrm{PA}}_i + \max_{j:(i,j)\in E} \mathrm{rank}^{\mathrm{PA}}_u(j)$.
%
This rank can be interpreted as an estimate of the remaining critical-path length when allowing partitions.

\textit{Partition-Aware Placement.}
The scheduler processes operators in decreasing order of partition-aware upward rank.
For each operator $i$, we evaluate each candidate partition plan $\pi_i$ and device assignment $\mathbf{d}_i$.
For a given pair $(\pi_i,\mathbf{d}_i)$, let $\mathrm{EST}(i,q,\pi_i,\mathbf{d}_i)$ denote the earliest feasible start time of subtask $q$, considering both predecessor completion times and current device availability.
Since active subtasks of the same operator are assigned to distinct devices, these start times can be computed independently without introducing intra-operator device conflicts.
The corresponding earliest finish time is $\mathrm{EFT}(i,\pi_i,\mathbf{d}_i) = \max_q \left[ \mathrm{EST}(i,q,\pi_i,\mathbf{d}_i) + \hat{\tau}_{i,a_i,d_{iq}}(u_{iq}) \right]$.
The heuristic greedily chooses partition plans and device assignment that minimize this earliest finish time: $(\pi_i^\star,\mathbf{d}_i^\star) = \arg\min_{\pi_i\in\Pi_i,\;\mathbf{d}_i\in\mathcal{D}^{K_i}:\\
d_{iq}\neq d_{ir}\forall q\neq r
} \mathrm{EFT}(i,\pi_i,\mathbf{d}_i)$.
This generalizes device selection in schedule-only list scheduling.
Instead of assigning an indivisible operator to one device, the heuristic selects both the workload division and the subtask-device placement for the current operator.

Compared with expanded-DAG scheduling, partition-aware list scheduling can account for current device availability and predecessor completion times when making a partition decision. 
However, the decision remains greedy; once an operator is scheduled, its partition plan and placement are fixed, and later decisions cannot revise it.
Thus, partition-aware list scheduling is a stronger practical baseline, but it still does not fully search the joint partition-scheduling space.

%% file: sections/5-method.tex
\section{Proposed Iterative Search}\label{sec:iterative_search}

The formulation in \cref{sec:baseline_formulation} captures both intra-operator parallelism through partition plans and inter-operator parallelism through DAG scheduling.
However, directly optimizing over all partition plans and schedules is too expensive.
We therefore propose an iterative search framework that improves the schedule through local updates to partition plans (\cref{sec:iterative_framework}).
In particular, three techniques facilitate a practical online search process.
First, staging decomposes a large inference DAG into smaller scheduling regions by exploiting block structure in neural networks (\cref{sec:iterative_staging}).
Second, criticality-aware sampling focuses candidate updates on operators that are likely to affect the makespan (\cref{sec:iterative_sampling}).
Third, latency prediction estimates the cost of partition plans without exhaustive on-device profiling (\cref{sec:iterative_latency_prediction}).

\subsection{Framework Overview}\label{sec:iterative_framework}

The full iterative search algorithm is summarized in the Appendix (\cref{sec:appendix:framework_details}, \cref{alg:iterative-search}).
The key idea is to separate partition-plan updates from schedule construction.
For each operator $i$, the search maintains a selected partition plan $\pi_i\in\Pi_i$, where $\pi_i=(a_i,\mathbf{u}_i)$ follows the notation in \cref{sec:baseline_formulation}.
We denote the partition plan at iteration $t$ by $\boldsymbol{\pi}^{(t)}=\{\pi_i^{(t)}\}_{i\in V}$.
Then, the schedule-construction algorithm builds a schedule $S^{(t)}=\textsc{Schedule}(G,\boldsymbol{\pi}^{(t)},\hat{\tau},\mathcal{D})$.
The produced schedule $S^{(t)}$ contains device assignments and execution order for the resulting subtasks; let $T(S^{(t)})$ denote its makespan.
Since \textsc{Schedule} receives fixed partition plans as input and does not need to solve the partitioning problem, it can be instantiated with an existing schedule-only heuristic, such as the list schedulers described in \cref{sec:baseline_schedule_only}.
The iterative search algorithm acts as the top-level search, which proposes partition-plan updates, uses \textsc{Schedule} to construct the corresponding schedule, and accepts an update that reduces the makespan.

At each iteration, the search algorithm first samples a set of partitionable operators; in this work, we propose the criticality-aware sampling rule described in \cref{sec:iterative_sampling}.
For each sampled operator $i$, the search algorithm generates a set of candidate updates;
a candidate update is denoted by $m=(i,\tilde{\pi}_i)$, where $\tilde{\pi}_i\in\Pi_i$ is a candidate replacement for the current partition plan $\pi_i^{(t)}$ of operator $i$.
In the default setting, the replacement plan $\tilde{\pi}_i$ is sampled by first choosing a partition strategy and then choosing a workload division uniformly at random.
Let $\boldsymbol{\pi}^{(t)}_m$ denote the partition-plan vector obtained after applying candidate update $m$ to the current plan vector $\boldsymbol{\pi}^{(t)}$: $\pi^{(t)}_{m,i}=\tilde{\pi}_i$ and $\pi^{(t)}_{m,j}=\pi_j^{(t)}$ for all $j\neq i$.
The corresponding candidate schedule is $S^{(t)}_m=\textsc{Schedule}(G,\boldsymbol{\pi}^{(t)}_m,\hat{\tau},\mathcal{D})$, and the makespan improvement is $\Delta^{(t)}_m=T(S^{(t)})-T(S^{(t)}_m)$.
Thus, $\Delta^{(t)}_m > 0$ means that the candidate update $m$ reduces the current makespan.
At each iteration, the algorithm accepts the candidate update with the largest positive improvement, and it stops when none of the sampled candidate updates improves the schedule.
The resulting procedure is a heuristic local search.
Although every accepted update reduces the estimated makespan, the use of sampled candidate updates means that the method provides neither a global optimality guarantee nor a worst-case approximation bound.
We therefore evaluate its solution quality empirically against offline joint optimization in \cref{sec:evaluation}.

\subsection{Staging}\label{sec:iterative_staging}

Neural networks are commonly organized as a sequence of blocks, where each block can contain parallel branches.
For example, an Inception block is a fork-join structure where branches are merged before the next block begins, as illustrated in \cref{fig:architecture_inceptionv3}.
In such structures, many scheduling decisions are local to a block, since later blocks cannot begin until the block output has been produced.
Staging exploits this observation by optimizing one region at a time instead of searching over the entire DAG at once.

A natural stage boundary is a \textit{global join point}.
We define a global join point as an operator that separates the DAG into a prefix and a suffix; i.e., every other operator is either an ancestor or a descendant of this operator.
Intuitively, when execution reaches such an operator, all computation in the prefix has joined, and the remaining suffix depends only on the joined output.
Closing a stage at a global join therefore gives a clean separation between the two regions.
Formally, let $\mathrm{Anc}(i)$ and $\mathrm{Desc}(i)$ denote the ancestors and descendants of operator $i$.
If $|\mathrm{Anc}(i)|+|\mathrm{Desc}(i)|=|V|-1$, then every other operator is comparable with $i$ in the DAG, and we treat $i$ as a global join point.
We use this criterion to identify stage boundaries when the model contains clear block-level joins.

However, some neural networks do not exhibit frequent global join points.
For example, HRNet (\cref{fig:architecture_hrnet}) contains long branch-heavy regions in which multiple branches progress independently before being merged much later.
If staging only depends on global joins, these regions can form large stages, making the search expensive.
To bound the search cost, we additionally enforce a maximum stage size (set to 20 operators in our experiments).

The way we explore the DAG determines which operators are included in the bounded stage: since iterative search constructs schedules using only the operators inside the current stage, 
the DAG exploration affects the schedule if a global join is not encountered before reaching the size limit.
In a branch-heavy region with multiple parallel branches, a naive exploration order may follow one branch too deeply, causing the bounded stage to contain mostly a dependent sequence of operators from the same branch, thus
exposing little inter-operator parallelism, resulting in a search that relies too heavily on intra-operator partitioning, potentially missing opportunities to schedule operators from different branches concurrently.
To avoid this, we explore the DAG according to decreasing HEFT upward rank $\mathrm{rank}_u(i)$, which estimates the remaining distance from operator $i$ to the graph exit, as defined in \cref{sec:baseline_schedule_only}.
Starting from the current stage, we repeatedly explore the available operator with the largest upward rank $\mathrm{rank}_u(i)$.
If this exploration reaches a global join point, the current stage is closed at that join.
If no global join is reached before the stage contains the maximum allowed number of operators, the stage is closed at the size limit.
This rank-guided exploration tends to select operators from parallel branches with similar remaining latency, rather than following a single branch for many consecutive operators.
As a result, bounded stages are more likely to preserve useful inter-operator parallelism.

After the DAG is divided into a list of stages, iterative search is run separately on each stage.
For each stage, predecessors that were scheduled in earlier stages provide completion times for operators in the current stage.
This preserves correctness with respect to the original dependencies, but can potentially restrict some cross-stage scheduling flexibility (in the case of reaching maximum stage size).
Thus, staging substantially reduces search cost while retaining most local scheduling opportunities inside neural-network blocks.
%
%
Without staging, searching over the entire model can hurt the optimization process and lead to a lower-quality schedule, as shown in the ablation study in \cref{sec:evaluation_ablation}.
%
%

\subsection{Criticality-Aware Sampling}\label{sec:iterative_sampling}

The main cost of the iterative search is update evaluation.
Each candidate update changes the partition plan of one operator and requires rebuilding the schedule to evaluate the makespan improvement.
Therefore, the search should spend most of its budget on operators whose partition plans can affect the current makespan.
Our sampling policy is motivated by a standard idea in critical-path scheduling: tasks on the critical path are more important because reducing their execution time can directly reduce the final completion time.
Specifically, we use \textit{slack} to quantify how close a scheduled subtask is to the critical path~\cite{kelley1959critical}.
%
%
Slack is the amount of time by which a scheduled subtask can be delayed without increasing the makespan.
A task with zero slack lies on at least one critical path of the realized schedule, while a task with large slack can move to a later point in the schedule without affecting end-to-end latency.
Therefore, changing a low-slack operator is more likely to reduce the makespan than changing a high-slack operator.

Because the current schedule may contain partitioned operators, we compute slack on the scheduled subtasks rather than on the original DAG.
Given the current schedule $S$, we construct a scheduled-subtask graph $\mathcal{G}_S = (\mathcal{V}_S, \mathcal{E}_S)$, where each node $(i,q) \in \mathcal{V}_S$ is a scheduled subtask with duration $\ell_{iq}=e_{iq}-s_{iq}$.
The edge set $\mathcal{E}_S$ contains two types of constraints: (1) dependency edges representing the full-barrier dependencies over subtasks from the original DAG, and (2) device-order edges connecting consecutive subtasks executed on the same device.
Criticality-aware sampling is then performed on $\mathcal{G}_S$.

We compute slack through a reverse pass over $\mathcal{G}_{S}$.
Let $T(S)$ be the makespan of the current schedule.
For each scheduled subtask $(i, q)$, let $LF_{iq}$ be the latest finish time that does not increase $T(S)$, and let $LS_{iq}$ be the corresponding latest start time.
If $(i,q)$ has no successor in $\mathcal{G}_S$, then $LF_{iq}=T(S)$;
otherwise, $LF_{iq}=\min_{(i,q)\rightarrow(j,r)\in\mathcal{E}_S} LS_{jr}$.
The latest start time is $LS_{iq}=LF_{iq}-\ell_{iq}$, and the resulting subtask slack is therefore $\sigma_{iq}=LS_{iq}-s_{iq}=LF_{iq}-e_{iq}$.
We aggregate subtask slack into an operator-level score by taking the minimum slack over the operator's subtasks: $\sigma_i=\min_{q:u_{iq}>0}\sigma_{i,q}$.
Thus, an operator is treated as important when any of its scheduled subtasks lie on (or close to) a tight path in the current schedule.

The sampler should focus on low-slack operators, but it should not sample only the current critical path.
Changing a non-critical operator can free a device earlier.
We therefore use a mixture distribution that combines criticality-biased sampling with uniform sampling.
We assign each operator a weight $\omega_i=\frac{\sum_{q} \ell_{iq}}{\sigma_i+\epsilon}$ as operators with longer duration (i.e., total subtask durations $\sum_{q} \ell_{iq}$) are more likely to have a greater impact on the makespan;
a small constant $\epsilon$ prevents division by zero.
The combined distribution is $P(i)=\rho\cdot \frac{\omega_i}{\sum_{j\in V_p} \omega_j} + (1-\rho)\cdot \frac{1}{|V_p|}$,
where the parameter $\rho$ controls the choice between two sampling approaches.
In our experiments, we use $\rho=0.8$ and $\epsilon=10^{-12}$.

Building $\mathcal{G}_S$ and computing slack are linear in the number of scheduled subtasks and schedule-order edges.
This cost is paid once per iteration and is small compared to update evaluation, which reruns the scheduling process for every sampled operator and candidate partition plan.

\subsection{Latency Prediction for Candidate Partition Plans}\label{sec:iterative_latency_prediction}

The iterative search repeatedly evaluates candidate partition plans $\pi_i=(a_i,\mathbf{u}_i)$, requiring latency estimation of a subtask with workload size $\mathbf{u}$ on device $d$ under partition strategy $a$.
Exhaustively profiling all such configurations during model initialization is impractical.
For example, our search on Inception-v3 considers 2462 operator configurations; measuring all of them on device would add substantial overhead.
Simple proxy metrics like FLOPs are insufficient because they do not capture memory behavior or framework-specific implementation choices~\cite{li2024inference,tang2021bridge}.
This limitation is especially substantial on GPUs, as GPU latency can change discontinuously with workload size because the backend may change the selected kernel implementation or map the workload to different GPU workgroup configurations~\cite{li2025epew}.
Consequently, a simple linear model over workload size cannot accurately predict partition latency.
On the other hand, complex ML predictors (e.g., MLP) can introduce too much latency prediction overhead for online search.

To balance prediction accuracy and runtime overhead, we adopt a gradient-boosted decision tree (GBDT) model \cite{friedman2001greedy}.
For each target device, we train (offline) separate GBDT predictors for each operator type (e.g., convolution, linear and element-wise operators); these predictors can be reused across neural networks.
Given operator $i$, partition strategy $a$, assigned workload size $u$, and device $d$, the predictor estimates the latency $\hat{\tau}_{i,a,d}(u)$.
The predictor input includes operator-level configuration features such as input and output tensor shapes, assigned output channels, kernel size, stride, padding, and FLOPs.
For GPU predictors, we further add backend-aware dispatch features following~\cite{li2025epew}, such as the GPU workgroup size and number of dispatched workgroups, which expose backend decisions that are hidden from operator configuration but substantially affect GPU execution time.
The complete feature table is summarized in the Appendix (\cref{sec:appendix:framework_details}, \cref{table:appendix_training_features}).
The resulting GBDT predictors (compiled into C code) are sufficiently efficient for online use.
For instance, for Inception-v3, the average prediction cost is 18.2~$\mu$s per operator configuration; evaluating all 2462 candidate configurations across three devices takes 135~ms in total, which is only 12.9\% of the 1048~ms model initialization time, as shown in \cref{sec:evaluation_overhead}.
The predictors provide sufficient accuracy for partition selection; e.g., convolution operators account for 90\% of the model's end-to-end GPU latency, and the corresponding mean absolute percentage error (MAPE) is below 8.2\% across all three devices.

%% file: sections/6-evaluation.tex
\section{Evaluation}\label{sec:evaluation}

In this section, we evaluate the proposed iterative search against baseline methods in both simulation (\cref{sec:evaluation_simulation}) and on-device execution (\cref{sec:evaluation_real_measurement}). We also evaluate its deployment-time overhead (\cref{sec:evaluation_overhead}) and conduct additional ablation and sensitivity analysis (\cref{sec:evaluation_ablation}).

\subsection{Experimental Setup}\label{sec:evaluation_setup}

\textit{Mobile Platforms.}
We evaluate $4$ mobile platforms, summarized in \cref{table:hardware_setup},
each modeled as three logical scheduling devices: GPU, CPU (L) for the large-core cluster, and CPU (M) for the medium-core cluster.
For CPU execution, threads are pinned within the selected cluster so that each measured latency corresponds to one logical device.
Considering that mobile measurements are sensitive to thermal throttling and dynamic frequency scaling, we follow prior benchmarking practice~\cite{li2024benchmark} to improve measurement stability by enabling performance mode when available, keeping the phone charging, and attaching an external cooling fan.

\begin{table}[t]
\centering
\scriptsize 
\begin{tabular}{lccc}
\toprule
\textbf{Family} & \textbf{\# Models} & \textbf{\# Convolution Range} & \textbf{\# Operators Range} \\
\midrule
Inception / Inception-ResNet & 4 & 94--244 & 170--429 \\
SqueezeNet / SqueezeResNet & 4 & 26 & 46--51 \\
PeleeNet & 1 & 113 & 144 \\
HRNet & 9 & 91--325 & 149--535 \\
\bottomrule
\end{tabular}
\caption{Workload summary. The full per-model list is reported in the Appendix (\cref{table:appendix_model_list}). }\label{table:workload_summary}
\end{table}

\textit{Workloads.}
We evaluate 18 image-classification models implemented from \textit{imgclsmob}~\cite{imgclsmob}; each model is exported to TensorFlow Lite with input image resolution $224{\times}224$ with batch size of 1.
\cref{table:workload_summary} summarizes the workload families, including Inception~\cite{szegedy2016rethinking}, Inception-ResNet~\cite{SzegedyIVA17}, SqueezeNet~\cite{iandola2016squeezenet}, PeleeNet~\cite{wang2018pelee}, and HRNet~\cite{wang2020deep}.
These models contain branch-heavy modules that expose substantial inter-operator parallelism, while their latency remains dominated by convolution operators for which intra-operator partitioning can be beneficial.
This diversity allows us to evaluate whether a scheduler can make effective decisions about both partitioning and scheduling.

\textit{On-Device Runtime.}
Our evaluation system has three components: latency predictors, an iterative scheduler, and an execution runtime.
First, the latency predictors estimate the execution time of candidate operator configurations.
We constructed a training dataset containing 10,000 samples of convolution configurations, 1,000 samples each for linear and element-wise (add/multiply) operators, and 500 samples for other operators, including padding, resizing, pooling, and activations.
A 20\% subset of the data was used for validation during training.
We train gradient-boosted decision tree (GBDT) models using LightGBM~\cite{lightgbm}, with hyperparameter settings following~\cite{li2025epew}.
The trained predictors are compiled into C code using Treelite~\cite{cho2018treelite} and TL2cgen~\cite{TL2cgen}, enabling efficient prediction on mobile CPUs.
This avoids exhaustively measuring every operator-device-partition configuration during model initialization.
Second, the iterative scheduler selects partitioning and scheduling decisions.
It is implemented in C++ and compiled into a native Android binary with the \texttt{-O3} optimization flag.
Unless otherwise stated, each stage contains at most 20 nodes; at each iteration, the search samples 5 candidate nodes and 10 candidate partitions per node.
Third, the execution runtime runs the generated schedule for end-to-end measurement, which incorporates CPU and GPU kernels from TensorFlow Lite v2.17.0;
specifically, GPU execution uses the OpenCL-based TensorFlow Lite GPU delegate~\cite{lee2019device}, while CPU execution uses XNNPACK kernels~\cite{xnnpack}.

\textit{Desktop Offline Evaluation.}
We use a desktop machine with an Intel Core i7-14700K CPU and 64~GB of memory for offline simulation and solver-based results (i.e., Gurobi).
All simulations are implemented in Python.
The Gurobi results are implemented with \texttt{gurobipy}~\cite{gurobi} and run with 28 CPU threads, with a timeout of 300~s for each stage obtained by dividing the original DAG.
These solver-based results are used to evaluate schedule quality, not as deployment-time scheduling methods.

\subsection{Simulation Results: Scheduling Quality}\label{sec:evaluation_simulation}

\begin{table}[t]
\centering
\scriptsize
\resizebox{\textwidth}{!}{%
\begin{tabular}{llrrrrrrrr}
\toprule
\multirow{2}{*}{Approach} & \multirow{2}{*}{Variant} & \multicolumn{2}{c}{OnePlus 11} & \multicolumn{2}{c}{Motorola 2022} & \multicolumn{2}{c}{Pixel 4} & \multicolumn{2}{c}{Pixel 5} \\
\cmidrule(lr){3-4}\cmidrule(lr){5-6}\cmidrule(lr){7-8}\cmidrule(lr){9-10}
 &  & Avg & Worst & Avg & Worst & Avg & Worst & Avg & Worst \\
\midrule
\multirow{3}{*}{Single-device} & GPU & 1.64 & 2.35 & 1.61 & 2.23 & 2.10 & 2.94 & 2.13 & 2.44 \\
 & CPU (L) & 7.57 & 10.05 & 7.71 & 9.46 & 6.31 & 7.27 & 4.87 & 6.30 \\
 & CPU (M) & 6.83 & 8.89 & 6.39 & 7.45 & 3.81 & 4.33 & 5.16 & 6.51 \\
\midrule
\multirow{1}{*}{Partition-only} & Min-local-latency & 1.20 & 1.39 & 1.22 & 1.59 & 1.16 & 1.42 & 1.11 & 1.28 \\
\midrule
\multirow{2}{*}{Schedule-only} & Best & 1.28 & 1.44 & 1.30 & 1.47 & 1.47 & 1.78 & 1.51 & 1.84 \\
 & Gurobi & 1.22 & 1.42 & 1.24 & 1.45 & 1.44 & 1.77 & 1.49 & 1.84 \\
\midrule
\multirow{4}{*}{Expanded-DAG} & Min-local-latency Best & 1.15 & 1.26 & 1.15 & 1.33 & 1.11 & 1.25 & 1.08 & 1.22 \\
 & Min-local-latency Gurobi & 1.09 & 1.15 & 1.08 & 1.22 & 1.05 & 1.15 & 1.04 & 1.13 \\
 & Equal-size Best & 1.34 & 1.52 & 1.34 & 1.60 & 1.25 & 1.42 & 1.18 & 1.36 \\
 & Equal-size Gurobi & 1.30 & 1.45 & 1.28 & 1.51 & 1.20 & 1.36 & 1.14 & 1.32 \\
\midrule
\multirow{1}{*}{Partition-aware} & HEFT & 1.12 & 1.21 & 1.12 & 1.29 & 1.08 & 1.19 & 1.07 & 1.18 \\
\midrule
\multirow{1}{*}{Iterative-search} & HEFT & 1.04 & 1.08 & 1.03 & 1.11 & 1.01 & 1.06 & 1.02 & 1.05 \\
\midrule
\end{tabular}%
}
\caption{Main simulation results. Latency is normalized to \textsc{Joint-Gurobi}. \textit{Best} denotes the best result among evaluated list-scheduling heuristics within each baseline method, while \textit{Gurobi} denotes the offline solver-based result for the corresponding restricted search space. \textit{Avg} (\textit{Worst}) reports the average (worst) normalized latency across 18 models. Full results are reported in the Appendix (\cref{table:appendix_simulation_full}).}
\label{table:eval_simulation_summary}
\end{table}

\begin{figure}[t]
    \centering
    \includegraphics[width=.9\linewidth]{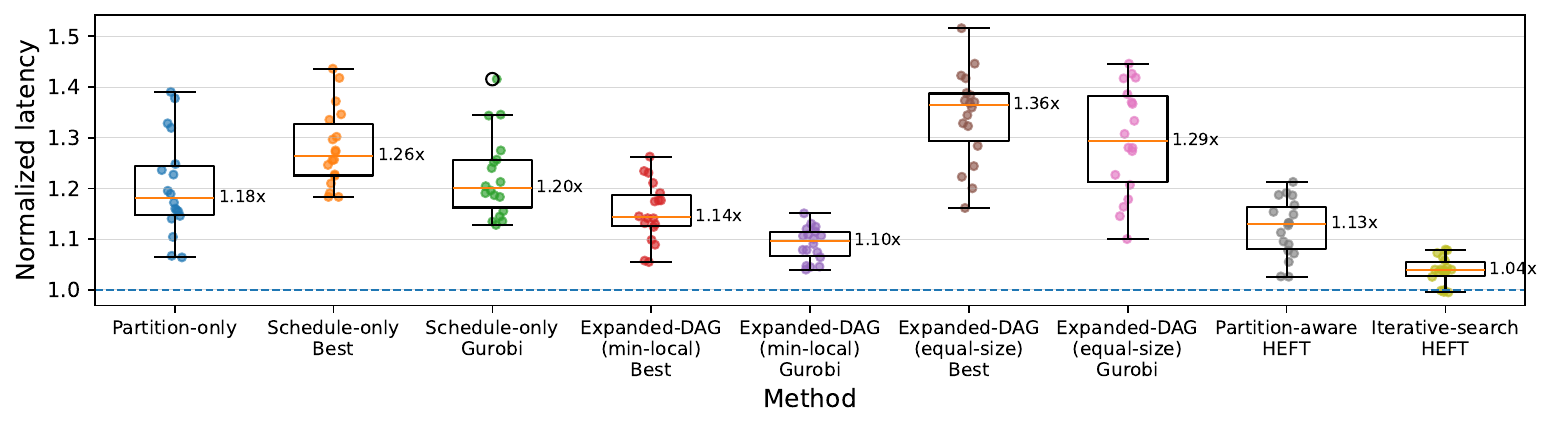}
    \caption{Distribution of normalized simulation latency across 18 models on OnePlus 11 (annotated values report the median). Latency is normalized to \textsc{Joint-Gurobi}. Additional results on other devices are reported in the Appendix (\cref{fig:appendix_simulation_full}).}
    \label{fig:evaluation_simulation_boxplot}
\end{figure}

We first evaluate scheduling quality in simulation using predicted operator-level latencies.
For each platform, latency is \textit{normalized} by the \textit{offline} result achieved by \textsc{Joint-Gurobi}, which solves the joint partition-aware scheduling formulation in \cref{sec:formulation}.
Notably, some approaches have multiple variants of schedule-construction algorithms.
For these approaches, \textit{Best} reports the lowest latency among the evaluated list-scheduling heuristics, and \textit{Gurobi} reports the corresponding offline solver-based result for that \textit{restricted} search space.
\cref{table:eval_simulation_summary} shows that single-device execution leaves substantial performance potential unexploited;
GPU is the fastest device, but it is still $1.61{\times}$--$2.13{\times}$ slower than \textsc{Joint-Gurobi} on average across platforms.
In addition, neither partition-only nor schedule-only execution is sufficient.
Partition-only execution improves significantly over GPU-only execution, reducing average normalized latency to $1.11{\times}$--$1.22{\times}$ across devices.
However, it remains slower than \textsc{Joint-Gurobi} because it chooses each partition using local operator latency rather than end-to-end DAG makespan.
Schedule-only methods also exhibit significant limitations due to unpartitioned operators.
Even schedule-only Gurobi is $1.22{\times}$--$1.49{\times}$ slower than \textsc{Joint-Gurobi} on average, which shows that the performance gap is not simply due to weak list-scheduling heuristics.

The expanded-DAG scheduling results show that exposing subtask partitioning is helpful but not sufficient.
With min-local-latency expansion, the best list scheduler reduces average normalized latency to $1.08{\times}$--$1.15{\times}$, and offline Gurobi scheduling of the expanded graph further reduces it to $1.04{\times}$--$1.09{\times}$.
However, these methods still fix the partition plan before scheduling, so they can miss globally better partition choices.
Equal-size expansion performs much worse, with average normalized latency between $1.18{\times}$ and $1.34{\times}$ for the best scheduling heuristic.
Partition-aware HEFT improves over expanded-DAG with best heuristics by choosing partition plans during scheduling, reaching $1.07{\times}$--$1.12{\times}$ average normalized latency.
However, it remains limited because its greedy local scheduling decisions can prevent the discovery of globally better schedules.

The proposed iterative search closes most of the gap to \textsc{Joint-Gurobi}.
Across the four devices, it achieves $1.01{\times}$--$1.04{\times}$ average normalized latency and $1.05{\times}$--$1.11{\times}$ worst-case normalized latency.
\cref{fig:evaluation_simulation_boxplot} depicts the distribution of normalized latency across models for each method on OnePlus 11; iterative search achieves a median normalized latency of $1.04{\times}$ and 90th-percentile latency of $1.07{\times}$, which further confirms that the improvement of iterative search is consistent across models.

\subsection{Real Measurements: End-to-End Latency}\label{sec:evaluation_real_measurement}

\begin{table}[t]
\centering
\scriptsize
\resizebox{\textwidth}{!}{%
\begin{tabular}{llrrrrrrrr}
\toprule
\multirow{2}{*}{Approach} & \multirow{2}{*}{Variant} & \multicolumn{2}{c}{OnePlus 11} & \multicolumn{2}{c}{Motorola 2022} & \multicolumn{2}{c}{Pixel 4} & \multicolumn{2}{c}{Pixel 5} \\
\cmidrule(lr){3-4}\cmidrule(lr){5-6}\cmidrule(lr){7-8}\cmidrule(lr){9-10}
 &  & Avg & Worst & Avg & Worst & Avg & Worst & Avg & Worst \\
\midrule
\multirow{3}{*}{Single-device} & GPU & 1.62 & 1.95 & 1.60 & 2.01 & 1.94 & 2.46 & 1.93 & 2.10 \\
 & CPU (L) & 7.68 & 11.72 & 7.45 & 10.29 & 6.60 & 9.11 & 4.45 & 5.54 \\
 & CPU (M) & 5.49 & 8.64 & 5.21 & 7.32 & 3.21 & 3.75 & 4.71 & 5.76 \\
\midrule
\multirow{1}{*}{Partition-only} & Min-local-latency & 1.37 & 2.92 & 1.25 & 1.61 & 1.20 & 1.93 & 1.12 & 1.52 \\
\midrule
\multirow{2}{*}{Schedule-only} & Best & 1.15 & 1.33 & 1.08 & 1.26 & 1.25 & 1.52 & 1.33 & 1.66 \\
 & Gurobi & 1.08 & 1.22 & 1.07 & 1.25 & 1.24 & 1.48 & 1.31 & 1.68 \\
\midrule
\multirow{4}{*}{Expanded-DAG} & Min-local-latency Best & 1.34 & 1.61 & 1.16 & 1.31 & 1.13 & 1.73 & 1.08 & 1.46 \\
 & Min-local-latency Gurobi & 1.19 & 1.33 & 1.13 & 1.36 & 1.09 & 1.69 & 1.07 & 1.53 \\
 & Equal-size Best & 1.68 & 2.08 & 1.43 & 1.80 & 1.22 & 1.90 & 1.20 & 1.82 \\
 & Equal-size Gurobi & 1.56 & 1.83 & 1.63 & 2.16 & 1.20 & 1.88 & 1.18 & 1.88 \\
\midrule
\multirow{1}{*}{Partition-aware} & HEFT & 1.36 & 1.69 & 1.18 & 1.39 & 1.11 & 1.78 & 1.10 & 1.60 \\
\midrule
\multirow{1}{*}{Iterative-search} & HEFT & 1.02 & 1.11 & 1.01 & 1.10 & 1.00 & 1.29 & 1.04 & 1.15 \\
\midrule
\end{tabular}%
}
\caption{Main measurement results. Latency is normalized to \textsc{Joint-Gurobi}. Full results are reported in the Appendix (\cref{table:appendix_measure_full}).}
\label{table:eval_measure_summary}
\end{table}

\begin{figure}[t]
    \centering
    \includegraphics[width=.6\linewidth]{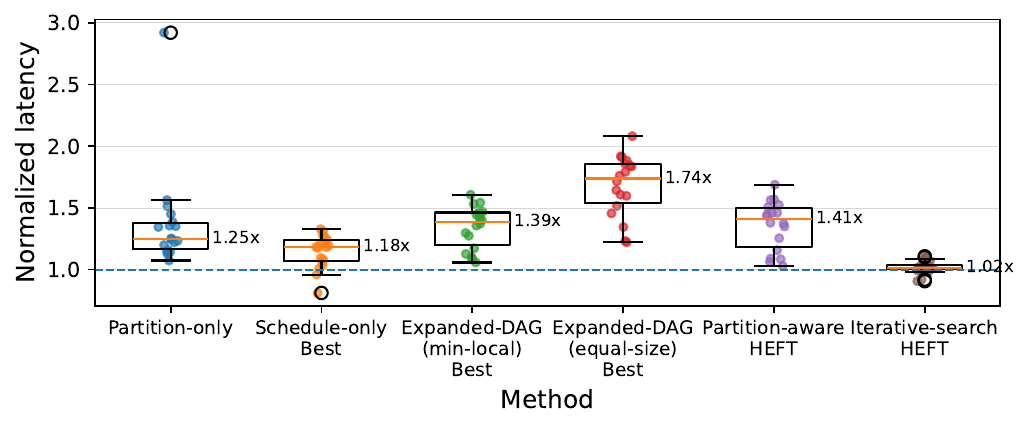}
    \caption{Distribution of measured end-to-end latency across 18 models on OnePlus 11, normalized to \textsc{Joint-Gurobi} (annotated values report the median). Additional results on other devices are reported in the Appendix (\cref{fig:appendix_measure_full}).}
    \label{fig:eval_real_measurement}
\end{figure}

We next evaluate the generated schedules on the mobile platforms.
Unlike the simulation study, this experiment executes the schedules through our runtime implementation and accounts for practical runtime effects such as kernel dispatch and CPU-GPU synchronization.
\cref{table:eval_measure_summary} shows that the main trends from simulation continue to hold on real devices, but with a clearer distinction between methods.
Single-device execution remains far from the joint optimization solution.
Schedule-only execution performs relatively well on some platforms because it dispatches fewer kernels and avoids partitioning overhead.
However, even offline schedule-only Gurobi remains $1.07{\times}$--$1.31{\times}$ slower than \textsc{Joint-Gurobi} on average, with worst-case ranging from $1.22{\times}$ to $1.68{\times}$.
This confirms that indivisible scheduling still misses intra-operator parallelism.
Partition-only execution is less consistent in real measurements.
Its average normalized latency ranges from $1.12{\times}$ on Pixel 5 to $1.37{\times}$ on OnePlus 11, and its worst case reaches $2.92{\times}$ on OnePlus 11.
This result suggests that excessive partitioning may deliver only marginal improvement because of dispatch and synchronization overhead in practice.
Iterative search is less sensitive to such overhead because it starts from a non-partitioning plan, accepting a partition update only if it reduces the overall makespan; thus, it can avoid excessive partitioning when inter-operator parallelism is available, facilitating utilization of devices by other ready operators.
The expanded-DAG methods show a similar effect; e.g., equal-size expansion performs poorly, reaching $1.68{\times}$ average normalized latency on OnePlus 11.
Even min-local-latency expanded-DAG scheduling can be worse than schedule-only execution on some devices, indicating that a partition plan that is locally effective may still be ineffective once runtime overhead and graph-level device availability are accounted for.

Iterative search is the most robust practical method.
Across the four phones, it achieves an average normalized latency within $1.00{\times}$--$1.04{\times}$ of \textsc{Joint-Gurobi}, with worst-case latency ranging from $1.10{\times}$ to $1.29{\times}$.
The distribution across models on OnePlus 11 (\cref{fig:eval_real_measurement}) further confirms this conclusion, where iterative search has a median normalized latency of $1.02{\times}$ and a 90th-percentile latency of $1.09{\times}$.
Overall, the real-device results show that iterative search preserves most of the benefit of joint partition-aware scheduling while avoiding the limitations of excessive partitioning, fixed expanded-DAG construction, or purely greedy partition-aware decisions.

\subsection{Runtime Overhead}\label{sec:evaluation_overhead}

\begin{figure}[t]
    \centering
    \includegraphics[width=.85\linewidth]{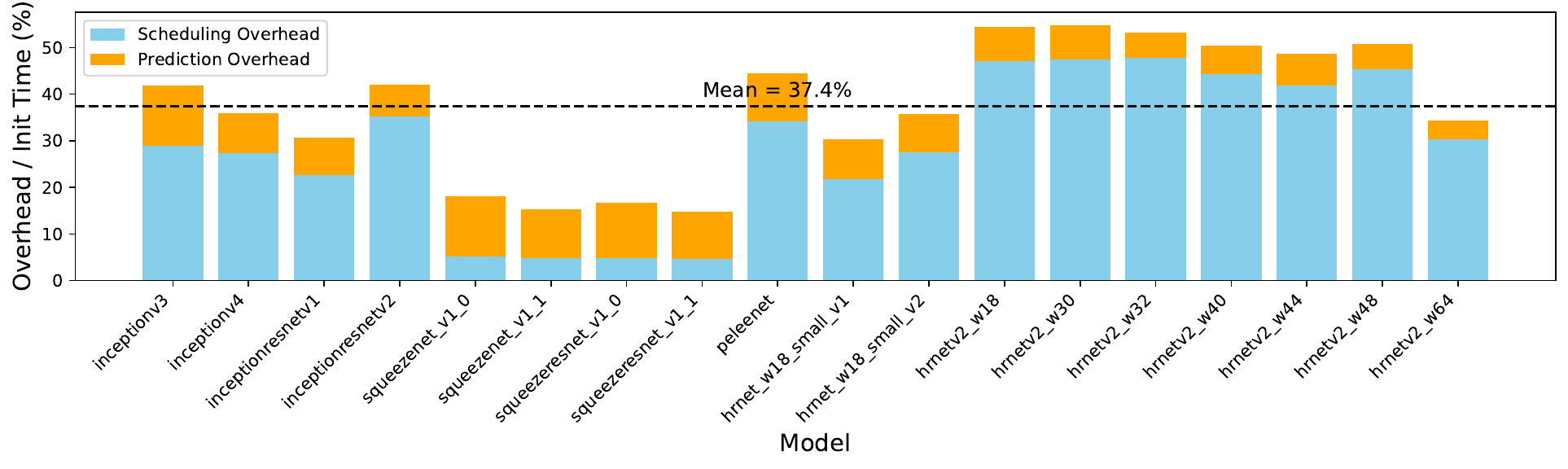}
    \caption{Scheduling overhead compared to TFLite model initialization time, measured on OnePlus 11. Each bar is stacked by prediction overhead and iterative scheduling overhead. Full measurements are reported in the Appendix (\cref{table:appendix_scheduling_overhead_full}).}\label{fig:eval_overhead}
\end{figure}

We next evaluate whether iterative search is practical to use during the model initialization process.
As discussed in \cref{sec:background_budget}, the schedule is computed once for a fixed model and input shape, cached, and reused across inference requests.
Therefore, the question is whether the overhead due to the iterative search is reasonable with respect to model initialization time rather than with respect to single-inference latency.
\cref{fig:eval_overhead} reports the runtime overhead on OnePlus 11.
We separate two components: latency-prediction overhead (in orange), which constructs the latency table for candidate partition plans, and scheduling overhead (in blue), which is incurred by the iterative search.

The latency-prediction overhead is relatively stable across models, ranging from 51~ms to 185~ms (4.0\% to 12.9\% of model initialization time).
Many configurations repeat within the same network, allowing predicted latencies to be reused.
Thus the scheduler can predict latency of candidate partitions without exhaustively measuring each operator-device-partition combination.
Scheduling overhead scales more directly with the number of partitionable operators and the size of the search space; it ranges from 24~ms for small SqueezeNet models to up to 1339~ms for the large HRNet models.
Consequently, scheduling accounts for a larger fraction of the total overhead of large models.
At the same time, model initialization time also grows with graph size and model complexity due to graph construction and memory allocation.
The total overhead is 37.4\% of model initialization time on average across all models.
For the larger models (e.g., HRNets), the overhead is mostly dictated by characteristics of the model family.
Regardless of model family, the total overhead is much smaller than exhaustive on-device profiling or solving the global optimization problem with an offline solver, which can take hours or days.

\subsection{Ablation and Sensitivity Analysis}\label{sec:evaluation_ablation}

\begin{figure}[t]
	\centering
	\begin{subfigure}[b]{.59\linewidth}
		\centering
		\includegraphics[width=\linewidth]{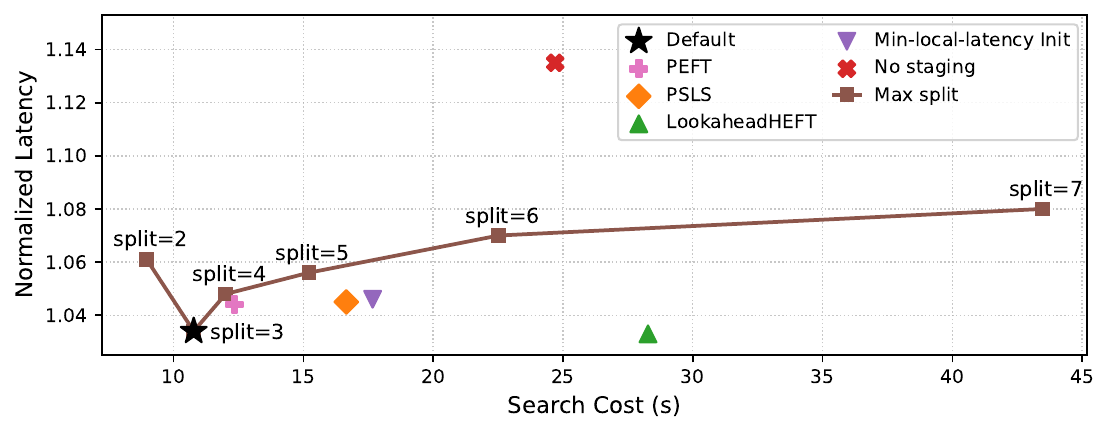}
		\caption{Ablation and split sensitivity. Numbers next to square markers denote the split value. The default setting uses staged search, HEFT schedule construction, and split of 3.}\label{fig:ablation_split_tradeoff}
	\end{subfigure}
	\begin{subfigure}[b]{.39\linewidth}
		\centering
		\includegraphics[width=\linewidth]{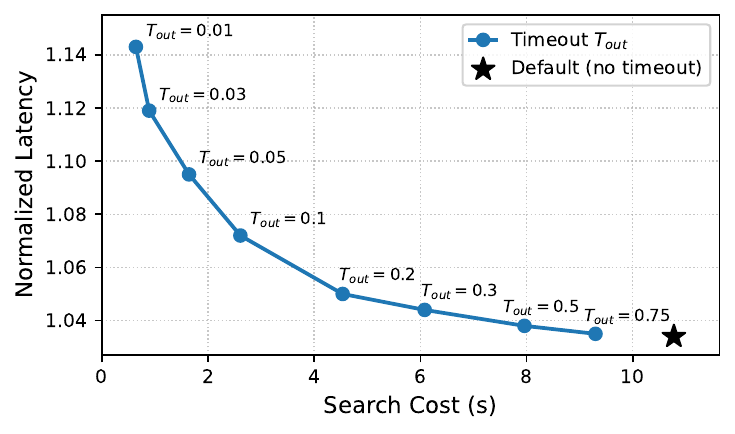}
		\caption{Timeout sensitivity. Smaller timeout values reduce search cost but degrade latency quality. The default setting has no timeout constraint.}\label{fig:timeout_tradeoff}
	\end{subfigure}
	\caption{Sensitivity of scheduling quality to iterative-search cost. Latency is normalized to \textsc{Joint-Gurobi}. Each point corresponds to one configuration, where lower normalized latency and lower search cost are preferred. Full data are reported in the Appendix (\cref{table:appendix_ablation_split}).}\label{fig:evaluation_ablation}
\end{figure}

The iterative search contains several design choices that trade scheduling quality for search overhead.
We study these choices to understand which components are most important and how the method behaves under different search budgets.
In this subsection, search cost is measured using the Python implementation on the desktop platform described in \cref{sec:evaluation_setup}.
The absolute search costs are therefore not directly comparable to the on-device overhead in \cref{sec:evaluation_overhead}, but the relative trends are useful for comparing design choices.
\cref{fig:evaluation_ablation} summarizes the trade-off between normalized latency (on a OnePlus 11 phone) and search cost, as detailed next.

\textit{Staging.}
Staging decomposes a large inference DAG into smaller scheduling regions.
Without staging, the search has a broader global view, but evaluating each candidate update becomes substantially more expensive, thus performing worse than the staged approach;
search cost increases from 10.8~s to 24.7~s, while normalized latency also increases from 1.03 to 1.14, under a 30-second search limit (which is already above the highest per-model search cost observed with staging).
Although the unstaged search can consider more of the graph at once, its higher per-update cost means that it evaluates fewer useful alternatives before the time limit.
This result suggests that staging is important not only for reducing search overhead but also for guiding the search toward useful local improvements.

\textit{Number of Splits.}
By default, we set the maximum number of subtasks per operator (i.e., splits) to the number of devices (\textit{three} in our experiments).
As shown in~\cref{fig:ablation_split_tradeoff}, this choice provides the best trade-off.
Using fewer splits reduces search cost but degrades schedule quality due to limited flexibility; e.g., split of 2 reduces search cost to 9.0~s but increases normalized latency to 1.06.
Using more splits does not improve quality and instead increases search cost, since each GPU kernel incurs a dispatch overhead, which may outweigh the benefits of additional flexibility due to finer-granularity partitions;
e.g., split of 7 raises the search cost to 43.5~s while producing worse normalized latency than split of 3.
This supports our default choice of setting the maximum split count to the number of devices.

\textit{Initialization.}
We compare two initial partition plans before search commences: no partitioning and min-local-latency partitioning.
The min-local-latency initialization achieves similar normalized latency but with increased search cost, from 10.8~s to 17.7~s.
This suggests that the final schedule quality is relatively robust to the initialization strategy, and that a locally optimal initialization does not necessarily lead to a better global schedule.

\textit{Scheduling Algorithms.}
We also evaluate alternative scheduling heuristics within the iterative-search framework, including PEFT, PSLS, and Lookahead-HEFT.
PEFT and PSLS produce slightly worse normalized latency than the HEFT-based default while increasing search cost.
Lookahead-HEFT achieves similar normalized latency, but requires 28.3~s of search, more than $2.6\times$ the default HEFT search cost.
Thus, HEFT provides the best practical trade-off as a schedule-construction algorithm.

\textit{Search Budget.}
Lastly, we vary the search budget by enforcing a timeout on the search for each stage.
As shown in~\cref{fig:timeout_tradeoff}, reducing the timeout produces a smooth trade-off between search cost and scheduling quality.
For example, a timeout of 0.3~s reduces search cost from 10.8~s to 6.1~s while maintaining similar normalized latency.
A more aggressive timeout of 0.2~s further reduces search cost to 4.5~s with normalized latency of 1.05.
Very small timeout values continue to reduce search cost, but the latency degradation becomes more significant.

\begin{table}[t]
\centering\small
\begin{tabular}{lcccc}
\toprule
\diagbox{\textbf{Scheduling}}{\textbf{Execution}} & \textbf{OnePlus 11} & \textbf{Motorola 2022} & \textbf{Pixel 4} & \textbf{Pixel 5} \\
\midrule
\textbf{OnePlus 11} & 1.02 & 1.06 & 1.23 & 1.27 \\
\textbf{Motorola 2022}  & 1.05 & 1.01 & 1.17 & 1.21 \\
\textbf{Pixel 4}    & 1.34 & 1.31 & 1.00 & 1.13 \\
\textbf{Pixel 5}    & 1.41 & 1.31 & 1.36 & 1.04 \\
\bottomrule
\end{tabular}
\caption{Cross-platform scheduling sensitivity. Each entry reports normalized latency when the schedule is generated using the row device and executed on the column device.}
\label{table:cross_platform}
\end{table}

\textit{Cross-Platform Scheduling.}
We now consider whether a schedule generated for one mobile platform can be reused on another platform.
In this experiment, the scheduler constructs an execution plan using latency predictions from one mobile platform, and the fixed plan is then executed on another platform.
The results in \cref{table:cross_platform} show that cross-platform reuse works well only when devices have similar CPU vs GPU performance characteristics.
For example, OnePlus 11 and Motorola 2022 transfer well in both directions, with up to $5\%$ increase in latency compared to performing the search on the actual device,
because both are recent Snapdragon 8-series platforms with similar CPU-GPU relative performance.
Further evidence of such similar performance characteristics is given in  
\cref{table:eval_measure_summary}, where single-device (GPU or CPU) executions achieve comparable normalized latency across the pairs of devices.
In contrast, schedules transfer poorly between newer Snapdragon 8-series devices and older Pixel devices, reaching up to $1.41{\times}$ when a Pixel 5 schedule is executed on OnePlus 11.
Thus, cross-platform reuse is possible for closely matched devices, but accurate per-device latency predictions remain important, e.g., when CPU-GPU relative performance changes.

\subsection{Discussion and Threats to Validity}\label{evaluation:discussion}

Our evaluation demonstrates the effectiveness of partition-aware scheduling under the workload and hardware assumptions considered in this work.
We next discuss the scope of these assumptions and factors that may affect the generalizability of our results to other inference workloads and hardware platforms.

\paragraph{Dynamic Inference Graphs}
Our formulation targets fixed-shape inference workloads represented by a static operator DAG, for which an execution plan can be constructed once during model initialization and reused across inference requests.
Although our experimental evaluation focused on CNNs, fixed-shape attention-based models (including most ViTs~\cite{dosovitskiy2020image}) would fit the same DAG-level formulation if appropriate attention-specific partition strategies and latency models are provided.
However, most autoregressive  (e.g., LLMs) and dynamically routed Mixture-of-Experts (MoE) models would have computation and resource demands that vary across tokens or inputs, violating our assumption of a reusable fixed execution plan.
Supporting such workloads would require dynamic scheduling, which we leave to future work.

\paragraph{Dynamic Batching}
We focus on latency-oriented mobile inference and use batch size of 1 throughout our evaluation.
For a \textit{fixed} batch size greater than one, the formulation would still keep a static DAG, but the corresponding tensor shapes would need to be incorporated into the latency models and partition plans.
Dynamic batching across multiple requests, however, introduces additional decisions, including when to form a batch and which requests to combine; this would also introduce throughput considerations in addition to single-request latency.
Joint batch formation and heterogeneous scheduling are outside the scope of the current formulation.

\paragraph{Accuracy of Latency Predictors}
Our scheduler relies on predicted operator latency to evaluate candidate partition plans; thus, prediction errors can affect partitioning and scheduling decisions.
The accuracy of operator latency prediction was evaluated in our earlier work on CNNs \cite{li2024inference,icpe23} and ViTs \cite{valuetools24}.
While we do not conduct a comprehensive sensitivity analysis of prediction error in this work, the cross-platform results in \cref{table:cross_platform} provide indirect evidence of its impact:
Schedules transfer well between platforms with similar CPU-GPU performance characteristics (1.05--1.06$\times$), but can degrade by up to 1.41$\times$ across dissimilar platforms.
This highlights the importance of accurately capturing relative device performance.

\paragraph{Communication Model}
Because our target mobile platforms provide unified CPU-GPU memory, we omit explicit tensor-copy communication costs while retaining synchronization and dispatch overhead.
This assumption does not directly hold for systems with non-uniform memory (e.g., discrete GPUs or distributed devices), where tensor copies would need to be incorporated into the cost model.

%% file: sections/7-related-work.tex
\section{Related Work}\label{sec:related_work}

\paragraph{Heterogeneous DAG Scheduling}
Static scheduling of DAGs on heterogeneous devices is a classical problem.
For instance, classical list-scheduling heuristics, including HEFT~\cite{topcuoglu2002performance}, Lookahead-HEFT~\cite{bittencourt2010dag}, PEFT~\cite{arabnejad2013list}, LDCP~\cite{daoud2008high}, CEFT~\cite{khan2012scheduling} and PSLS~\cite{zhao2019list}, assign each DAG task to one device while respecting dependency constraints.
These methods differ in how they rank tasks, estimate downstream criticality, or select devices.
However, they share the same task abstraction, where each DAG node is an indivisible unit that must execute on one device.
This abstraction is appropriate for many task-graph workloads, but it misses an important opportunity in mobile DNN inference:
Latency-dominant operators such as convolutions can often be partitioned across CPU and GPU, so treating every operator as an atomic task can leave devices idle in chain-dominated models.
Our work extends the scheduling decision space by allowing selected operators to be split into subtasks whose placement and timing are optimized together with the DAG schedule.
At the same time, our iterative framework can reuse these list schedulers as scheduling modules once a candidate partition plan is fixed.

\paragraph{Meta-Heuristic, Learning-Based, and Offline Scheduling Approaches}
Beyond list scheduling, meta-heuristic and learning-based methods explore larger scheduling spaces.
Genetic algorithms~\cite{ding2017hybrid,sulaiman2021evolutionary}, particle-swarm methods~\cite{shirvani2020hybrid}, and chemical-reaction optimization methods~\cite{xu2013dag} have been applied to heterogeneous DAG scheduling.
Learning-based systems use reinforcement learning~\cite{mirhoseini2017device,mao2019learning,bojja2019learning} or graph neural networks~\cite{zhou2022learning} to learn device-placement or scheduling policies.
These approaches show that richer search can outperform simple list scheduling, especially when workload structure is complex.
However, they typically require substantial offline training, repeated simulation, or many runtime evaluations, and they often target cloud clusters, distributed training graphs, or general DAG workloads.
Instead, our goal is to ensure that our scheduler runs during mobile model initialization and fits within a deployment-time budget. 
Therefore, we use offline optimization (with a long timeout of 5 minutes per stage) only as a notion of an optimal solution and design a lightweight iterative search that reduces the search space through staging, criticality-aware sampling, and latency prediction, while attempting to produce a solution close to that of the offline approach.

\paragraph{DNN Partitioning for Edge and Collaborative Inference}
Another line of work partitions DNN computation across distributed edge-cloud platforms.
Systems such as Neurosurgeon~\cite{kang2017neurosurgeon} and JointDNN~\cite{eshratifar2019jointdnn} study layer-level partitioning
between mobile devices and cloud servers, while DDNN~\cite{teerapittayanon2017distributed} and Edgent~\cite{li2019edge} consider hierarchical edge-cloud execution, early exits, or adaptive model sizing.
These systems primarily decide where \textit{coarse-grained} model subgraphs or distributed inference tasks should run across networked platforms.
In that setting, network bandwidth, feature-map transfer, and offloading latency are central optimization factors.
In contrast, our work targets execution on heterogeneous hardware units within a single mobile device.
The key challenge is not where to offload a subgraph across the network, but how to jointly exploit intra-operator partitioning and inter-operator DAG scheduling across local CPU and GPU resources.

\paragraph{Mobile Heterogeneous DNN Co-Execution}
Recent systems study fine-grained CPU-GPU co-execution for mobile inference.
CoDL~\cite{jia2022codl} shows that a DNN operator can be partitioned across mobile CPU and GPU to improve performance and energy efficiency.
Our previous work \cite{li2025epew} further improves fine-grained co-execution by reducing synchronization overhead and improving latency prediction for partitioned operators.
These works are closely related to ours because they demonstrate that mobile CPU and GPU can effectively cooperate within a single operator.
However, they mainly optimize partitioning at the operator level.
They do not address the full DAG scheduling problem, where multiple operators may be ready at the same time and where using all devices for one partitioned operator can delay other ready operators.
Our work extends this line of research from isolated operator co-execution to end-to-end inference DAG execution.
Our results show that the key question is not only how to partition an operator, but also whether that operator should be partitioned in the current graph context.

\paragraph{Joint Partitioning and Scheduling}
The closest work is HeSP~\cite{rey2016hesp}, which studies a scheduling-partitioning problem, incorporating recursive task partitioning as an additional degree of freedom alongside scheduling.
Their idea is closely related to our motivation, since both works observe that partitioning and scheduling should not be optimized independently; partitioning changes the task graph seen by the scheduler, while scheduling determines whether the extra parallelism introduced by partitioning is actually useful.
However, HeSP is designed as an (offline) \textit{simulation} framework and is evaluated mainly on dense linear-algebra task graphs with GPU servers and CPU-based edge devices. In contrast, our approach targets mobile neural-network inference DAGs, where our scheduler must be sufficiently efficient to be used online during model initialization, rather than as an offline simulation or solver. Given the neural-network setting, we also introduce partitioning strategy (i.e., operator-level workload splits, such as output-channel, input-channel, or spatial partitioning) as an additional search dimension.
Given our online (model initialization time) focus, in contrast to \cite{rey2016hesp}, we design staging, criticality-aware sampling, and latency prediction to capture most of the benefit of joint optimization but within a limited deployment-time budget.

%% file: sections/8-conclusion.tex
\section{Conclusion}

Our work shows that efficient mobile inference on devices with heterogeneous accelerators requires jointly considering operator partitioning and DAG scheduling.
Schedule-only methods can exploit inter-operator parallelism, but often leave devices underutilized, particularly in chain-dominated models.
Partition-only methods improve utilization by greedily co-executing operators across CPU and GPU through intra-operator parallelism, but they commonly miss opportunities to utilize an available device for other ready operators.
To exploit both forms of parallelism, we formulate this problem as partition-aware DAG scheduling, which jointly considers partition choices, workload division, device assignment, and execution order.
To make our solution practical for deployment time, we design an iterative search framework that uses staged optimization, criticality-aware sampling, and lightweight latency prediction to reduce optimization overhead.
As a result, the proposed scheduler achieves latency close to that of offline joint optimization -- across representative models and mobile platforms -- while keeping overhead within a practical model-initialization budget.
These results demonstrate that combining intra-operator co-execution with inter-operator DAG scheduling is an effective approach for accelerating mobile inference.
Future directions include extending the framework to specialized on-device accelerators such as NPUs and DSPs, and incorporating additional objectives such as energy consumption and thermal behavior.

\section*{Acknowledgments}

\noindent This work was supported in part by the NSF CNS-1816887, CCF-1763747, and IIS-1833137 awards.

%% file: sections/appendix.tex
\appendix\section*{Appendix}

\section{Runtime Implementation for Scheduled Co-Execution}\label{sec:appendix:runtime_details}
\label{sec:appendix:synchronization_details}

This appendix describes how the execution plan produced by our scheduler is executed at runtime.
After partitioning decisions are fixed, we view the computation as an expanded-DAG.
Specifically, each node represents either an original operator or one partitioned piece of an operator, where edges represent dependencies based on the full-barrier dependencies from the original DAG.
The scheduler assigns each node to a device and determines the execution order of nodes on each device.
The runtime execution follows this plan by maintaining one ordered queue per device.
Each device executes its own queue in the order chosen by the scheduler.

The runtime execution maintains DAG dependencies using a shared dependency state.
For each node, the runtime execution records how many of its predecessor nodes are still unfinished; a node becomes ready when this count reaches zero.
When a node finishes, the runtime execution updates the dependency states of its successor nodes.
This simple mechanism also captures the full-barrier dependency model used by the scheduler; i.e., if an operator is partitioned, any successor can start only after all required predecessor pieces have completed.

The CPU implements this dependency directly.
After a CPU node finishes, the CPU updates the shared dependency states of its successors using lightweight atomic operations.
Before starting the next node in its queue, the CPU checks whether that node is ready and waits only if some predecessor is still unfinished.
In our measurements, this CPU-side synchronization overhead is negligible compared with operator execution time.

On the other hand, the GPU requires an additional step because GPU work is submitted asynchronously through a command queue.
The CPU can enqueue GPU computation kernels in the scheduled order, but dependency updates that occur after a GPU computation must also respect this GPU queue order.
Therefore, when a GPU node needs to publish its completion or wait for work produced by CPU, the runtime execution inserts a small GPU-side synchronization kernel.
This kernel only updates or checks the shared dependency state and is inserted after each GPU operator computation kernel.
Although these GPU-side synchronization kernels perform little computation, launching a GPU kernel still incurs dispatch overhead because the CPU must submit the kernel to the GPU command queue.
Accordingly, our schedule evaluator accounts for the overhead of dispatching a synchronization kernel by adding a small constant to GPU execution time.
In our implementation, this constant is set to $10~\mu$s, based on the average dispatch overhead measured in our experiments.

\section{Supplemental Framework Details}\label{sec:appendix:framework_details}

\begin{algorithm}[t]
\caption{Iterative Search for Partition-Aware Scheduling}
\label{alg:iterative-search}
\textbf{Inputs:} search budget $B$, operator DAG $G=(V,E)$, devices $\mathcal{D}$, candidate plans $\{\Pi_i\}_{i\in V}$, \\\hspace*{5mm} partitionable operators $V_p = \{i \in V : |\Pi_i| > 1\}$, latency predictor $\hat{\tau}$, scheduling routine \pr{Schedule}, \\\hspace*{5mm} operator sampler \pr{SampleOperators}, update generator \pr{GenerateUpdates}\\
\textbf{Outputs:} Final partition-plan assignment and schedule 
\begin{pseudo}[indent-mark,indent-mark-shift]
     $\boldsymbol{\pi} \gets  \text{initial partition } \{\pi_i\}_{i\in V}$ \label{line:init-plan} \\
    $S\gets \pr{Schedule}(G,\boldsymbol{\pi},\hat{\tau},\mathcal{D})$ \label{line:init-schedule} \\
    \kw{for} $t=1,\ldots,B$ \\+
        $V_{\mathrm{samp}}\gets \pr{SampleOperators}(G,V_p,\boldsymbol{\pi},S)$ \label{line:sample-ops} \\
        $\boldsymbol{\pi}^\star\gets \boldsymbol{\pi}$, $S^\star\gets S$, $\Delta^\star\gets 0$ \label{line:track-best-start} \\
        \kw{for} each $i\in V_{\mathrm{samp}}$ \\+
            $\mathcal{M}_i\gets \pr{GenerateUpdates}(i,\Pi_i,\pi_i)$ \label{line:generate-moves} \\
            \kw{for} each $m=(i,\tilde{\pi}_i)\in\mathcal{M}_i$ \\+
                Construct $\boldsymbol{\pi}_m$ by replacing $\pi_i$ with $\tilde{\pi}_i$ \\
                $S_m\gets \pr{Schedule}(G,\boldsymbol{\pi}_m,\hat{\tau},\mathcal{D})$ \label{line:evaluate-move} \\
                $\Delta_m\gets T(S)-T(S_m)$ \label{line:update-delta}\\
                \kw{if} $\Delta_m>\Delta^\star$ \\+
                     $\boldsymbol{\pi}^\star\gets \boldsymbol{\pi}_m$, $S^\star\gets S_m$, $\Delta^\star\gets \Delta_m$ \label{line:track-best-end} \\---
        \kw{if} $\Delta^\star=0$  \\+
            \kw{return} $\boldsymbol{\pi}, S$ \label{line:terminate}\\-
        $\boldsymbol{\pi}\gets \boldsymbol{\pi}^\star$, 
        $S\gets S^\star$ \label{line:accept-plan}\label{line:accept-schedule} \\-
    \kw{return} $\boldsymbol{\pi}, S$
\end{pseudo}
\end{algorithm}

\begin{table}[t]
\centering
\small
\renewcommand{\arraystretch}{1.1}
\begin{tabular}{p{1.8cm} p{1.8cm} p{10.5cm}}
\toprule
\textbf{Operator} & \textbf{Device} & \textbf{Features} \\
\midrule

\multirow{2}{*}{Convolution}
& GPU & Input/output height, input/output width, input/output channels, filter shape, stride, group count, input/output sizes, filter size, FLOPs, grid size ($x,y,z$-dims), workgroup size ($x,y,z$-dims), workgroup count ($x,y,z$-dims), total workgroup count \\
\cmidrule(lr){2-3}
& CPU & Input/output height, input/output width, input/output channels, filter shape, stride, group count, input/output sizes, filter size, FLOPs \\
\midrule

\multirow{2}{*}{Linear}
& GPU & Input/output height, input/output width, input/output channels, input/output sizes, weight size, FLOPs, grid size ($x,y,z$-dims), workgroup size ($x,y,z$-dims), workgroup count ($x,y,z$-dims), total workgroup count \\
\cmidrule(lr){2-3}
& CPU & Input/output height, input/output width, input/output channels, input/output sizes, weight size, FLOPs \\
\midrule

Pooling & CPU / GPU & Input/output height, input/output width, input/output channels, input/output sizes, kernel shape, stride \\
\midrule

Others & CPU / GPU & Input/output height, input/output width, input/output channels, input/output sizes \\

\bottomrule
\end{tabular}
\caption{Features used for training GBDT predictors across different operators and devices.}
\label{table:appendix_training_features}
\end{table}

This appendix provides additional details of our iterative search, including (1) pseudo-code corresponding to our iterative search, and (2) a feature table for building our latency predictors.
\cref{alg:iterative-search} summarizes our iterative search framework.
Lines~\ref{line:init-plan}--\ref{line:init-schedule} initialize the partition plan (e.g., no partitioning in our default setting) and schedule.
Line~\ref{line:sample-ops} selects a subset of partitionable operators on which to spend the search budget of each iteration; our default implementation uses schedule slack to prioritize operators near the critical path, as described in \cref{sec:iterative_sampling}.
Lines~\ref{line:generate-moves}--\ref{line:evaluate-move} generate candidate replacement plans and reschedule the resulting graph.
Lines~\ref{line:update-delta}--\ref{line:track-best-end} keep the best improved update in the current iteration.
If no sampled update improves the makespan, line~\ref{line:terminate} terminates the search.
Otherwise, line~\ref{line:accept-plan} accepts the best candidate assignment and continues to the next iteration.
We set the search budget $B$ to a large value that it is never reached in our experiments; it serves only as a safeguard against unusually long searches, while termination is normally determined by the absence of improving candidate updates.

\cref{table:appendix_training_features} summarizes the operator, partition, and GPU dispatch features used by the latency predictors.

\section{Supplemental Data}\label{sec:appendix:supplementary_data}

\begin{table}[t]
\centering
\scriptsize 
\begin{tabular}{lrr}
    \toprule
    \textbf{Model} & \textbf{\# Conv} & \textbf{\# Ops} \\
    \midrule
    inceptionv3 & 94 & 170 \\
    inceptionv4 & 149 & 271 \\
    inceptionresnetv1 & 132 & 231 \\
    inceptionresnetv2 & 244 & 429 \\
    squeezenet\_v1\_0 & 26 & 47 \\
    squeezenet\_v1\_1 & 26 & 46 \\
    squeezeresnet\_v1\_0 & 26 & 51 \\
    squeezeresnet\_v1\_1 & 26 & 50 \\
    peleenet & 113 & 144 \\
    hrnet\_w18\_small\_v1 & 91 & 149 \\
    hrnet\_w18\_small\_v2 & 164 & 279 \\
    hrnetv2\_w18 & 325 & 535 \\
    hrnetv2\_w30 & 325 & 535 \\
    hrnetv2\_w32 & 325 & 535 \\
    hrnetv2\_w40 & 325 & 535 \\
    hrnetv2\_w44 & 325 & 535 \\
    hrnetv2\_w48 & 325 & 535 \\
    hrnetv2\_w64 & 325 & 535 \\
    \bottomrule
\end{tabular}
\caption{Neural-network workloads. All models are exported to TFLite and evaluated with input resolution $224{\times}224$.}\label{table:appendix_model_list}
\end{table}

\begin{table}[t]
\centering
\scriptsize
\setlength{\tabcolsep}{3pt}
\renewcommand{\arraystretch}{0.95}
\resizebox{\textwidth}{!}{%
\begin{tabular}{llrrrrrrrr}
\toprule
\multirow{2}{*}{Approach} & \multirow{2}{*}{Variant} & \multicolumn{2}{c}{OnePlus 11} & \multicolumn{2}{c}{Motorola 2022} & \multicolumn{2}{c}{Pixel 4} & \multicolumn{2}{c}{Pixel 5} \\
\cmidrule(lr){3-4}\cmidrule(lr){5-6}\cmidrule(lr){7-8}\cmidrule(lr){9-10}
 &  & Avg & Worst & Avg & Worst & Avg & Worst & Avg & Worst \\
\midrule
\multirow{1}{*}{Joint-search} & Gurobi & 1.00 & 1.00 & 1.00 & 1.00 & 1.00 & 1.00 & 1.00 & 1.00 \\
\midrule
\multirow{3}{*}{Single-device} & GPU & 1.64 & 2.35 & 1.61 & 2.23 & 2.10 & 2.94 & 2.13 & 2.44 \\
 & CPU (L) & 7.57 & 10.05 & 7.71 & 9.46 & 6.31 & 7.27 & 4.87 & 6.30 \\
 & CPU (M) & 6.83 & 8.89 & 6.39 & 7.45 & 3.81 & 4.33 & 5.16 & 6.51 \\
\midrule
\multirow{1}{*}{Partition-only} & Min-local-latency & 1.20 & 1.39 & 1.22 & 1.59 & 1.16 & 1.42 & 1.11 & 1.28 \\
\midrule
\multirow{7}{*}{Schedule-only} & HEFT & 1.29 & 1.42 & 1.31 & 1.47 & 1.47 & 1.78 & 1.51 & 1.84 \\
 & Lookahead-HEFT & 1.29 & 1.42 & 1.31 & 1.47 & 1.47 & 1.78 & 1.52 & 1.84 \\
 & PEFT & 1.29 & 1.45 & 1.33 & 1.56 & 1.49 & 1.82 & 1.53 & 1.92 \\
 & LDCP & 1.28 & 1.44 & 1.30 & 1.47 & 1.47 & 1.78 & 1.52 & 1.84 \\
 & CEFT & 1.49 & 1.61 & 1.50 & 1.68 & 1.66 & 2.08 & 1.69 & 2.07 \\
 & PSLS & 1.28 & 1.49 & 1.31 & 1.57 & 1.49 & 1.80 & 1.56 & 1.85 \\
 & Gurobi & 1.22 & 1.42 & 1.24 & 1.45 & 1.44 & 1.77 & 1.49 & 1.84 \\
\midrule
\multirow{7}{*}{Expanded-DAG (Min-local-latency)} & HEFT & 1.15 & 1.26 & 1.15 & 1.32 & 1.11 & 1.26 & 1.09 & 1.20 \\
 & Lookahead-HEFT & 1.15 & 1.27 & 1.15 & 1.32 & 1.11 & 1.25 & 1.09 & 1.20 \\
 & PEFT & 1.15 & 1.26 & 1.15 & 1.33 & 1.24 & 1.35 & 1.08 & 1.22 \\
 & LDCP & 1.16 & 1.27 & 1.16 & 1.36 & 1.11 & 1.25 & 1.09 & 1.22 \\
 & CEFT & 1.35 & 1.56 & 1.35 & 1.54 & 1.29 & 1.45 & 1.27 & 1.44 \\
 & PSLS & 1.17 & 1.33 & 1.17 & 1.36 & 1.11 & 1.23 & 1.09 & 1.21 \\
 & Gurobi & 1.09 & 1.15 & 1.08 & 1.22 & 1.05 & 1.15 & 1.04 & 1.13 \\
\midrule
\multirow{7}{*}{Expanded-DAG (Equal-size)} & HEFT & 1.34 & 1.52 & 1.34 & 1.60 & 1.25 & 1.42 & 1.18 & 1.36 \\
 & Lookahead-HEFT & 1.34 & 1.52 & 1.34 & 1.60 & 1.25 & 1.42 & 1.18 & 1.36 \\
 & PEFT & 1.40 & 1.55 & 1.38 & 1.59 & 1.30 & 1.44 & 1.22 & 1.39 \\
 & LDCP & 1.35 & 1.53 & 1.35 & 1.61 & 1.27 & 1.45 & 1.19 & 1.38 \\
 & CEFT & 1.50 & 1.70 & 1.48 & 1.80 & 1.35 & 1.55 & 1.28 & 1.50 \\
 & PSLS & 1.37 & 1.54 & 1.37 & 1.58 & 1.29 & 1.42 & 1.21 & 1.36 \\
 & Gurobi & 1.30 & 1.45 & 1.28 & 1.51 & 1.20 & 1.36 & 1.14 & 1.32 \\
\midrule
 \multirow{1}{*}{Partition-aware heuristic} & HEFT & 1.12 & 1.21 & 1.12 & 1.29 & 1.08 & 1.19 & 1.07 & 1.18 \\
\midrule
\multirow{4}{*}{Iterative search} & HEFT & 1.04 & 1.08 & 1.03 & 1.11 & 1.01 & 1.06 & 1.02 & 1.05 \\
 & PEFT & 1.05 & 1.12 & 1.04 & 1.12 & 1.03 & 1.08 & 1.03 & 1.07 \\
 & PSLS & 1.05 & 1.13 & 1.04 & 1.14 & 1.03 & 1.08 & 1.03 & 1.06 \\
 & Lookahead-HEFT & 1.04 & 1.08 & 1.03 & 1.12 & 1.01 & 1.06 & 1.02 & 1.05 \\
\bottomrule
\end{tabular}%
}
\caption{Simulation results across four phones. Latency is normalized to \textsc{Joint-Gurobi}; lower is better.}
\label{table:appendix_simulation_full}
\end{table}

\begin{figure}[t]
	\centering
	\begin{subfigure}[b]{.9\linewidth}
		\centering
		\includegraphics[width=\linewidth]{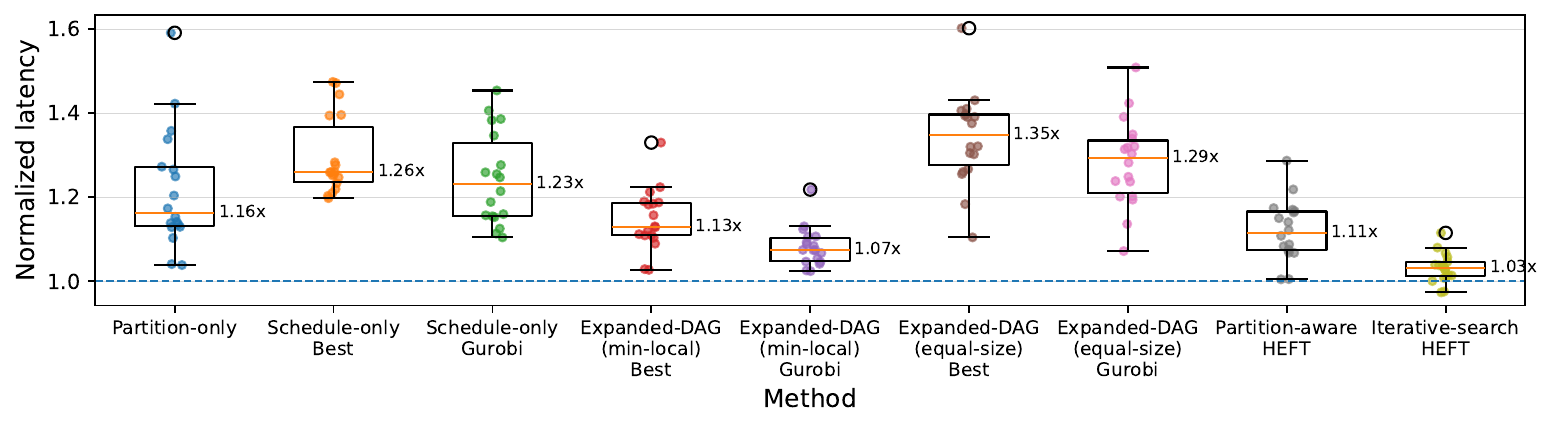}
		\caption{Motorola 2022}\label{fig:appendix_simulation_moto2022}
	\end{subfigure}
	\begin{subfigure}[b]{.9\linewidth}
		\centering
		\includegraphics[width=\linewidth]{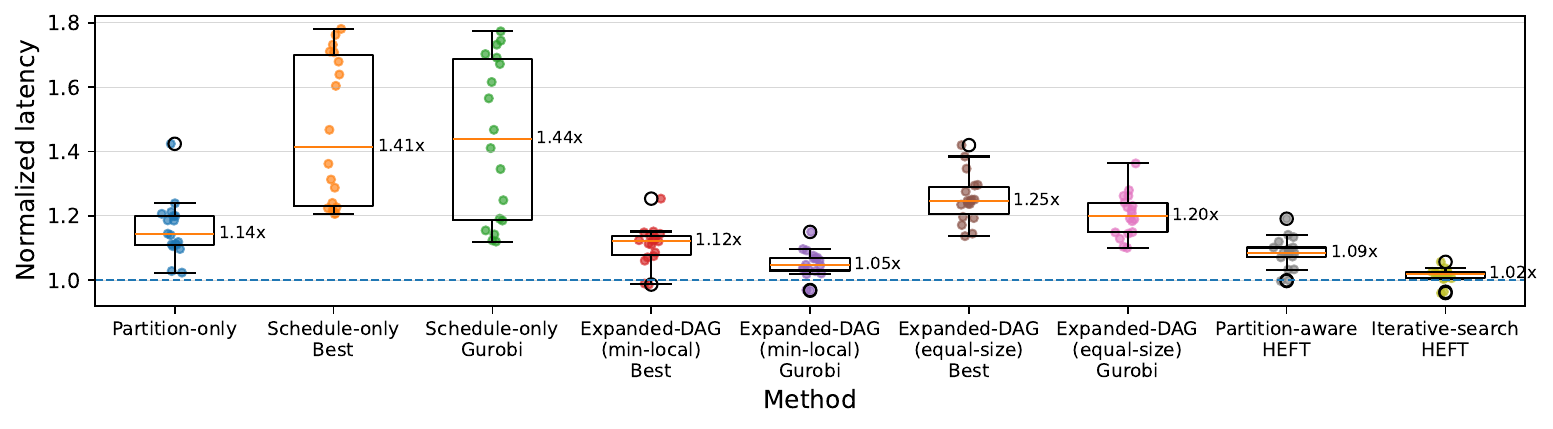}
		\caption{Pixel 4}\label{fig:appendix_simulation_pixel4}
	\end{subfigure}
	\begin{subfigure}[b]{.9\linewidth}
		\centering
		\includegraphics[width=\linewidth]{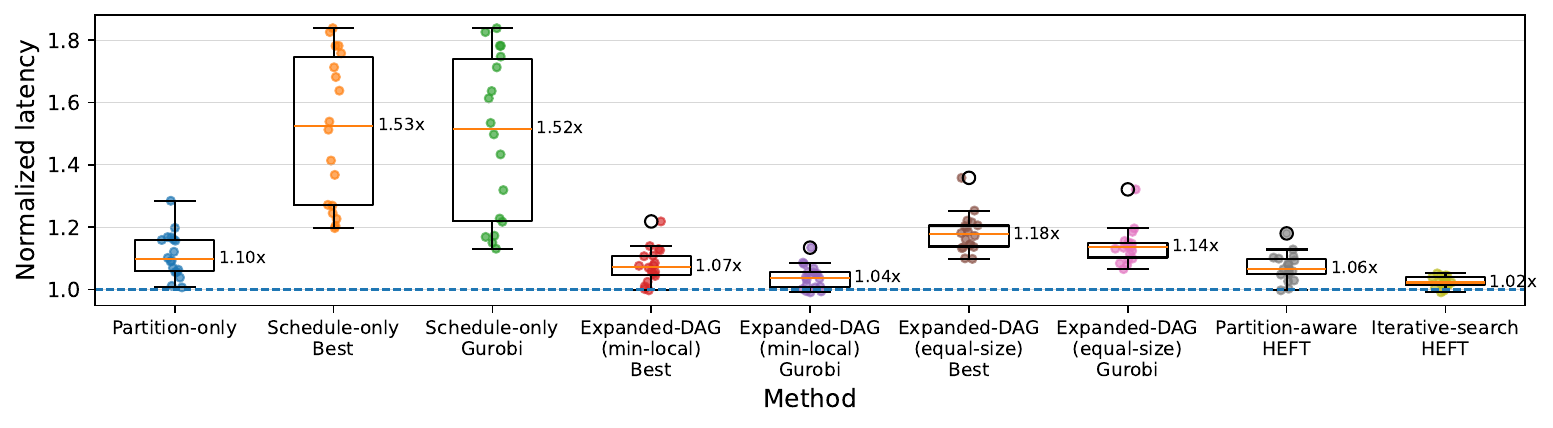}
		\caption{Pixel 5}\label{fig:appendix_simulation_pixel5}
	\end{subfigure}
	\caption{Distribution of simulation latency across 18 models, normalized to \textsc{Joint-Gurobi} (annotated values report the median)}\label{fig:appendix_simulation_full}
\end{figure}

\begin{table}[t]
\centering
\scriptsize
\setlength{\tabcolsep}{3pt}
\renewcommand{\arraystretch}{0.95}
\resizebox{\textwidth}{!}{%
\begin{tabular}{llrrrrrrrr}
\toprule
\multirow{2}{*}{Approach} & \multirow{2}{*}{Variant} & \multicolumn{2}{c}{OnePlus 11} & \multicolumn{2}{c}{Motorola 2022} & \multicolumn{2}{c}{Pixel 4} & \multicolumn{2}{c}{Pixel 5} \\
\cmidrule(lr){3-4}\cmidrule(lr){5-6}\cmidrule(lr){7-8}\cmidrule(lr){9-10}
 &  & Avg & Worst & Avg & Worst & Avg & Worst & Avg & Worst \\
\midrule
\multirow{1}{*}{Joint-search} & Gurobi & 1.00 & 1.00 & 1.00 & 1.00 & 1.00 & 1.00 & 1.00 & 1.00 \\
\midrule
\multirow{3}{*}{Single-device} & GPU & 1.62 & 1.95 & 1.60 & 2.01 & 1.94 & 2.46 & 1.93 & 2.10 \\
 & CPU (L) & 7.68 & 11.72 & 7.45 & 10.29 & 6.60 & 9.11 & 4.45 & 5.54 \\
 & CPU (M) & 5.49 & 8.64 & 5.21 & 7.32 & 3.21 & 3.75 & 4.71 & 5.76 \\
\midrule
\multirow{1}{*}{Partition-only} & Min-local-latency & 1.37 & 2.92 & 1.25 & 1.61 & 1.20 & 1.93 & 1.12 & 1.52 \\
\midrule
\multirow{7}{*}{Schedule-only} & HEFT & 1.16 & 1.73 & 1.08 & 1.30 & 1.26 & 1.45 & 1.33 & 1.66 \\
 & Lookahead-HEFT & 1.25 & 2.67 & 1.09 & 1.30 & 1.26 & 1.44 & 1.33 & 1.66 \\
 & PEFT & 1.25 & 1.73 & 1.12 & 1.31 & 1.25 & 1.52 & 1.36 & 1.76 \\
 & LDCP & 1.15 & 1.33 & 1.08 & 1.26 & 1.25 & 1.45 & 1.34 & 1.66 \\
 & CEFT & 1.33 & 1.50 & 1.27 & 1.52 & 1.39 & 1.67 & 1.49 & 2.00 \\
 & PSLS & 1.16 & 1.36 & 1.10 & 1.31 & 1.30 & 1.45 & 1.37 & 1.66 \\
 & Gurobi & 1.08 & 1.22 & 1.07 & 1.25 & 1.24 & 1.48 & 1.31 & 1.68 \\
\midrule
\multirow{7}{*}{Expanded-DAG (Min-local-latency)} & HEFT & 1.38 & 2.56 & 1.17 & 1.40 & 1.13 & 1.83 & 1.10 & 1.49 \\
 & Lookahead-HEFT & 1.39 & 2.60 & 1.17 & 1.40 & 1.13 & 1.86 & 1.10 & 1.50 \\
 & PEFT & 1.42 & 2.86 & 1.19 & 1.32 & 1.22 & 1.75 & 1.08 & 1.46 \\
 & LDCP & 1.34 & 1.61 & 1.16 & 1.31 & 1.13 & 1.86 & 1.11 & 1.69 \\
 & CEFT & 1.77 & 3.05 & 1.41 & 1.69 & 1.26 & 2.06 & 1.23 & 1.59 \\
 & PSLS & 1.37 & 1.95 & 1.21 & 1.45 & 1.13 & 1.73 & 1.11 & 1.44 \\
 & Gurobi & 1.19 & 1.33 & 1.13 & 1.36 & 1.09 & 1.69 & 1.07 & 1.53 \\
\midrule
\multirow{7}{*}{Expanded-DAG (Equal-size)} & HEFT & 1.73 & 2.50 & 1.44 & 1.81 & 1.22 & 1.90 & 1.20 & 1.82 \\
 & Lookahead-HEFT & 1.68 & 2.08 & 1.43 & 1.80 & 1.22 & 1.89 & 1.21 & 1.84 \\
 & PEFT & 1.85 & 2.24 & 1.50 & 1.82 & 1.26 & 1.94 & 1.22 & 1.84 \\
 & LDCP & 1.75 & 2.34 & 1.45 & 1.72 & 1.22 & 1.91 & 1.21 & 1.89 \\
 & CEFT & 1.86 & 2.29 & 1.57 & 2.01 & 1.28 & 1.83 & 1.26 & 1.91 \\
 & PSLS & 1.76 & 2.14 & 1.49 & 1.81 & 1.29 & 1.85 & 1.23 & 1.90 \\
 & Gurobi & 1.56 & 1.83 & 1.63 & 2.16 & 1.20 & 1.88 & 1.18 & 1.88 \\
\midrule
 \multirow{1}{*}{Partition-aware heuristic} & HEFT & 1.36 & 1.69 & 1.18 & 1.39 & 1.11 & 1.78 & 1.10 & 1.60 \\
\midrule
\multirow{4}{*}{Iterative search} & HEFT & 1.02 & 1.11 & 1.01 & 1.10 & 1.00 & 1.29 & 1.04 & 1.15 \\
 & PEFT & 1.12 & 1.44 & 1.07 & 1.23 & 1.00 & 1.57 & 1.08 & 1.42 \\
 & PSLS & 1.05 & 1.12 & 1.03 & 1.20 & 1.02 & 1.30 & 1.06 & 1.19 \\
 & Lookahead-HEFT & 1.06 & 1.54 & 1.02 & 1.13 & 1.01 & 1.21 & 1.04 & 1.13 \\
\bottomrule
\end{tabular}%
}
\caption{Full measurement results across four phones. Latency is normalized to \textsc{Joint-Gurobi}; lower is better}
\label{table:appendix_measure_full}
\end{table}

\begin{figure}[t]
	\centering
	\begin{subfigure}[b]{.49\linewidth}
		\centering
		\includegraphics[width=\linewidth]{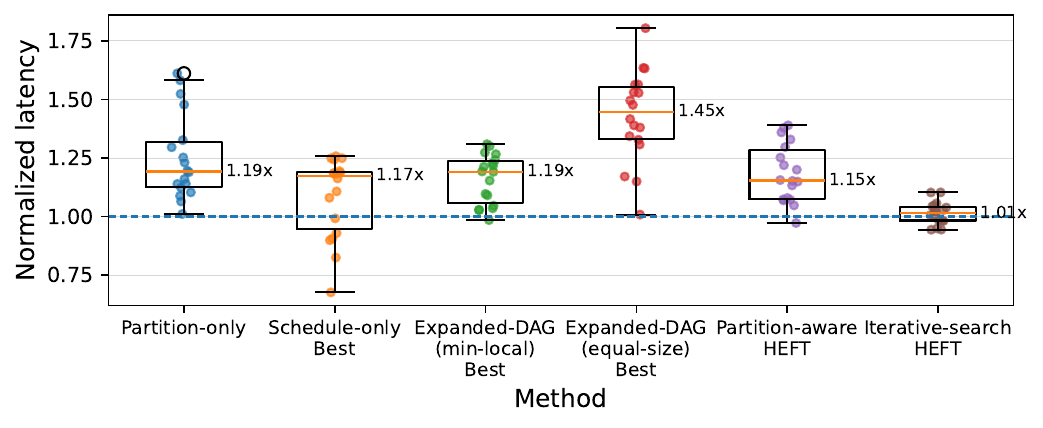}
		\caption{Motorola 2022}\label{fig:appendix_measure_moto2022}
	\end{subfigure}
	\begin{subfigure}[b]{.49\linewidth}
		\centering
		\includegraphics[width=\linewidth]{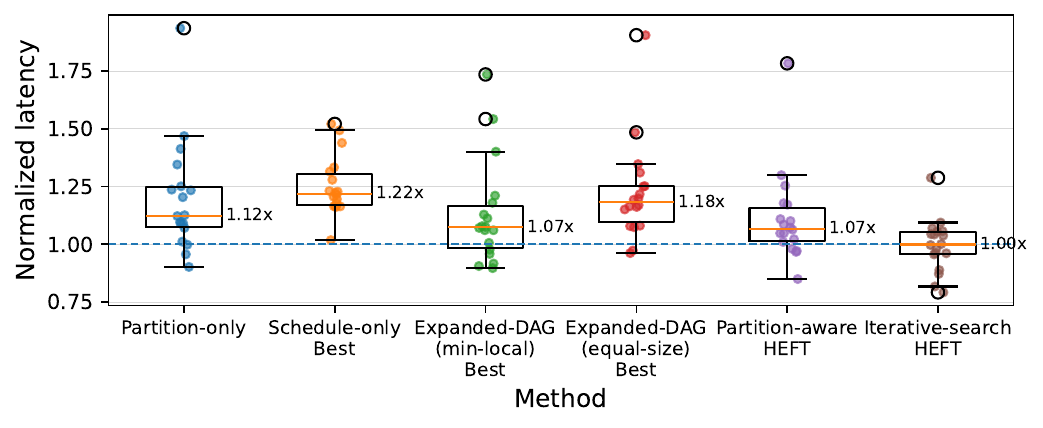}
		\caption{Pixel 4}\label{fig:appendix_measure_pixel4}
	\end{subfigure}
	\begin{subfigure}[b]{.49\linewidth}
		\centering
		\includegraphics[width=\linewidth]{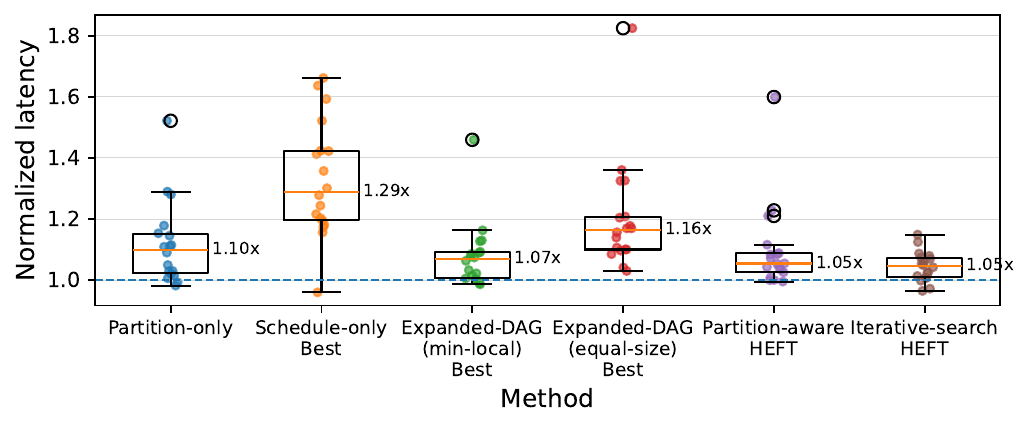}
		\caption{Pixel 5}\label{fig:appendix_measure_pixel5}
	\end{subfigure}
	\caption{Distribution of measured end-to-end latency across 18 models, normalized to \textsc{Joint-Gurobi}  (annotated values report the median)}\label{fig:appendix_measure_full}
\end{figure}

\begin{table}[t]
\centering
\scriptsize 
\begin{tabular}{lrrr}
\toprule
Model & Prediction overhead (ms) & Scheduling overhead (ms) & Initialization time (ms) \\
\midrule
inceptionv3 & 135 & 303 & 1048 \\
inceptionv4 & 130 & 420 & 1534 \\
inceptionresnetv1 & 88 & 249 & 1100 \\
inceptionresnetv2 & 123 & 627 & 1783 \\
squeezenet\_v1\_0 & 62 & 25 & 479 \\
squeezenet\_v1\_1 & 51 & 24 & 482 \\
squeezeresnet\_v1\_0 & 60 & 24 & 504 \\
squeezeresnet\_v1\_1 & 51 & 24 & 504 \\
peleenet & 81 & 265 & 778 \\
hrnet\_w18\_small\_v1 & 122 & 311 & 1424 \\
hrnet\_w18\_small\_v2 & 134 & 446 & 1625 \\
hrnetv2\_w18 & 134 & 855 & 1817 \\
hrnetv2\_w30 & 163 & 1046 & 2204 \\
hrnetv2\_w32 & 124 & 1084 & 2272 \\
hrnetv2\_w40 & 156 & 1140 & 2574 \\
hrnetv2\_w44 & 185 & 1127 & 2692 \\
hrnetv2\_w48 & 161 & 1339 & 2955 \\
hrnetv2\_w64 & 148 & 1116 & 3675 \\
\bottomrule
\end{tabular}
\caption{Overhead compared with TFLite model initialization time}\label{table:appendix_scheduling_overhead_full}
\end{table}

\begin{table}[t]
\centering
\scriptsize
\begin{tabular}{llcc}
\toprule
 & Configuration & Normalized latency & Search cost (s) \\
\midrule
\multirow{1}{*}{\textbf{Default}} & \textbf{HEFT, split=3, staging} & \textbf{1.03} & \textbf{10.8} \\
\midrule
\multirow{4}{*}{Heuristics} & PEFT & 1.04 & 12.3 \\
 & PSLS & 1.04 & 16.7 \\
 & Lookahead-HEFT & 1.03 & 28.3 \\
 & Min-local-latency Init & 1.05 & 17.7 \\
\midrule
\multirow{1}{*}{Staging} & no staging & 1.14 & 24.7 \\
\midrule
\multirow{7}{*}{Number of splits} & split=2 & 1.06 & 9.0 \\
 & split=3 & 1.03 & 10.8 \\
 & split=4 & 1.05 & 12.0 \\
 & split=5 & 1.06 & 15.2 \\
 & split=6 & 1.07 & 22.5 \\
 & split=7 & 1.08 & 43.5 \\
 & split=8 & 1.09 & 103.9 \\
\bottomrule
\end{tabular}
\caption{Ablation and split-sensitivity results for iterative search. Lower normalized latency (on OnePlus 11) and lower search cost are preferred.}
\label{table:appendix_ablation_split}
\end{table}

\begin{table}[t]
\centering
\scriptsize
\begin{tabular}{lcc}
\toprule
Timeout (s) & Normalized latency & Search cost (s) \\
\midrule
Unlimited & 1.03 & 10.78 \\
0.75 & 1.03 & 9.30 \\
0.5 & 1.04 & 7.96 \\
0.4 & 1.04 & 7.12 \\
0.3 & 1.04 & 6.08 \\
0.2 & 1.05 & 4.54 \\
0.1 & 1.07 & 2.61 \\
0.05 & 1.09 & 1.64 \\
0.04 & 1.09 & 1.50 \\
0.03 & 1.12 & 0.89 \\
0.02 & 1.13 & 0.75 \\
0.01 & 1.14 & 0.64 \\
\bottomrule
\end{tabular}
\caption{Timeout-sensitivity results for iterative search. Lower timeout values reduce search cost but can degrade normalized latency (measured on OnePlus 11).}
\label{table:appendix_timeout}
\end{table}

This appendix provides supplemental data for the information and results provided in the main text.
\cref{table:appendix_model_list} summarizes the full specifications of models in our study.
\cref{table:appendix_simulation_full} reports the complete simulated latency results for all baselines across the four mobile platforms.
\cref{fig:appendix_simulation_full} depicts the distributions of simulated latencies across the 18 models on the three platforms not included in the main text.
\cref{table:appendix_measure_full} reports the complete real-device latency measurements for all baselines across the four mobile platforms.
\cref{fig:appendix_measure_full} depicts the distributions of measured latencies across the 18 models on the three platforms not included in the main text.
\cref{table:appendix_scheduling_overhead_full} compares TensorFlow Lite model initialization time with the latency-prediction and iterative-scheduling overheads.
\cref{table:appendix_ablation_split,table:appendix_timeout} provide the complete ablation results for iterative-search design choices and search-budget sensitivity.